\documentclass[11pt,twoside]{article}
\usepackage{atmp}
\copyrightnotice{2002}{6}{141}{196}
\def\beq{\begin{equation}}
\def\eeq{\end{equation}}
\def\bea{\begin{eqnarray}}
\def\eea{\end{eqnarray}}
\def\bef{\begin{figure}}
\def\enf{\end{figure}}
\def\S{{\bf S}}
\def\A{{\bf A}}
\def\B{{\bf B}}
\def\C{{\bf C}}
\def\D{{\bf D}}
\def\E{{\bf E}}
\def\Z{{\bf Z}}
\def\R{{\bf R}}
\def\P{{\bf P}}
\def\N{{\bf N}}

\def\z{{\zeta}}
\def\RP{{\bf RP}}

\def\CF{{\cal F}}

\def\CI{{\cal I}}
\def\CN{{\cal N}}
\def\CO{{\cal O}}
\def\CR{{\cal R}}

\def\CT{{\cal T}}

\def\We{{W_{\mbox{eff}}}}
\def\Wd{{W_{\mbox{deg}}}}
\def\Wq{{W_{\mbox{q}}}}
\def\Tr{{\mbox{Tr}}}

\def\mjk{{M_{j,k}}}
\def\mlm{{M_{l,m}}}
\def\ljk{{\Lambda_{j,k}}}
\def\gjk{{g_{j,k}}}
\def\ajk{{a_{j,k}}}
\def\alm{{a_{l,m}}}

\def\li{{\Lambda_{i}}}

\def\ba{\begin{array}}
\def\ea{\end{array}}
\def\bce{\begin{center}}
\def\ece{\end{center}}
\def\non{{\nonumber}}

\def\ev#1{\langle#1\rangle}

\def\be{\beta}

\def\ep{\epsilon}

\def\La{\Lambda}

\begin{document}
\title
{Duality and Confinement in $\CN =1$ Supersymmetric
Theories from Geometric Transitions} 
\url{hep-th/0112040}
\vspace*{-.5in}

\author{Kyungho Oh}
\vspace*{-.3in}

\address{Department of Mathematics, Computer Science, Physics and Astronomy\\
University of Missouri-St. Louis\\
Saint Louis, MO 63121}
\addressemail{oh@arch.umsl.edu} 
\vspace*{-.5in}

\author{Radu Tatar}
\vspace*{-.3in}

\address{Institut fur Physik\\ Humboldt
University\\ Berlin, 10115, Germany}
\addressemail{tatar@physik.hu-berlin.de}
\markboth{\it Duality and Confinement in $\CN =1$ 
}{\it K. Oh and R. Tatar}

\begin{abstract} We study large N dualities for a general class of 
${\cal N} = 1$ theories realized on type IIB D5 branes wrapping 2-cycles of 
local
Calabi-Yau threefolds or as effective field theories on D4 branes in type IIA 
brane configurations. We completely
solve the issue of the classical moduli space for ${\cal N} = 2, 
\prod_{i=1}^n U(N_i)$ theories deformed by a
general superpotential for the adjoint and bifundamental fields. The 
$\CN=1$ geometries in type IIB and its
T-dual brane configurations are presented and they agree with the field 
theory analysis. We investigate the
geometric transitions in the ten dimensional theories as well as in M-theory. 
Strong coupling effects in field
theory are analyzed in the deformed geometry with fluxes. 
Gluino condensations are identified the normalizable
deformation parameters while the vacuum expectation values of the 
bifundamental fields are with the
non-normalizable ones. By lifting to M theory, we get a transition 
from finite coverings of non-hyperelliptic
curves to non-hyperelliptic curves. We also discuss orientifold theories, 
Seiberg dualities and mirror
symmetries.
\end{abstract}
\cutpage

\section{Introduction}
\pagestyle{myheadings}
\markboth{Duality and Confiement in $N=1$...}{K. Oh and R. Tatar}

Duality is one of the most fascinating aspects  in string theory. Recently, geometric transitions became an important
tool in understanding large N dualities between open string theory and closed string theory. In geometric
transitions,  one begins with  D-branes wrapped around cycles of a Calabi-Yau threefold and the field theory is
described in the small 't Hooft parameter region. After a geometric transition of the Calabi-Yau threefold to
another Calabi-Yau threefold, the D-branes disappear and they are replaced by RR and NS fluxes through the dual
cycles, the field theory being described in the large 't Hooft parameter region.

The large $N$ duality between Chern-Simons theory on the $\bf S^3$ cycle of the deformed conifold and topological
closed strings on the resolved conifold was observed in \cite{gova}  and the result was embedded in type IIA
strings by Vafa \cite{vafa}. The topological transition becomes a geometric transition between D6 branes on the
$\bf S^3$ of the deformed conifold and type IIA strings on the resolved conifold, with fluxes or between D5 branes
wrapped on the $\P^1$ cycle of the resolved conifold and type IIB on the deformed conifold with fluxes. The type
IIB formulation has been extended to a large class of geometric transition dualities, for geometries which are
more complicated than the conifold \cite{sv,civ,eot,ckv,GUD,dot1,dot2,Klebanov,KW}.

In the present work we explore the large N dualities from type IIB, type IIA and M-theory perspectives for large
classes of the $\CN=1$ supersymmetric gauge theories by wrapping D5 branes on blown-up $\P^1$ cycles in
Calabi-Yau threefolds which are obtained by deforming resolved ALE spaces of the ADE singularities. We begin with
the  ${\cal N} = 2$ quiver gauge associated with the ADE Dynkin diagrams, which can be geometrically engineered
on the resolved ALE spaces. Then the $\CN=1$ theory is obtained by  adding a superpotential for the adjoint
fields, $W_i(\Phi_i)$, thus the full tree-level  superpotential is \bea \label{SUPER} W =\sum_{i=1}^n W_i
-\mbox{Tr}\sum_{i=1}^{n}\sum_{j=1}^n s_{i,j}Q_{i,j} \Phi_{j} {Q}_{j,i},~~\mbox{where }  W_i =
\mbox{Tr}\sum_{j=1}^{d_i+1} \frac{g_{i,j-1}}{j} \Phi_i^j, \eea  with adjoints $\Phi$ and bifundamentals
$Q_{i,j}$, which fixes the moduli space of the  ${\cal N} = 1$ theory. This case has not been explicitly
discussed in the field theory literature and we provide full and detailed explanation on how the moduli space of
the  ${\cal N} = 1$ theory appears as a solution of the F-term and D-term equations from ${\cal N} = 2$ field
theory deformed by the superpotential (\ref{SUPER}). In the vacuum, the gauge group is broken as: \bea
\prod_{i=1}^n U(N_i) \rightarrow \prod_{j=1}^n \prod_{k=j}^n \prod_{l=1}^{d_{j,k}} U(M_{j,k,l}) \eea where
$d_{j,k}$ is the maximum of the degrees $W_i'$ for $j\leq i \leq k$.

Having identified the classical moduli, we interpret the expectation values of the adjoints and adjoints in terms
of complex and topological  structure changes in Calabi-Yau threefolds in type IIB picture, and correspondingly
curving of NS branes and fixing of D4 branes in type IIA picture.  Moreover, we identify the number of the Higgsed
gauge groups in the $\CN=1$ theory  with the number of the $\P^1$ cycles in type IIB and the number of
intersection points of NS branes in type IIA. These interpretations enable us to write down the $\CN=1$ geometry
explicitly, which is a small resolution of the singular threefold given by \bea \label{n1geometry} xy - u
\prod_{p=1}^n \left(u - \sum_{i=1}^p W'_i(v)\right) =0. \eea

The deformations of three dimensional ADE-type of singularities  can be divided into two classes in the sense of
\cite{Gukov}. In the $\CN=1$ theory, the non-normalizable deformations are fixed by the expectation values of the
bifundamentals which result in topological changes of the Calabi-Yau spaces
by creating $\S^3$ cycles in type IIB,  and in turn
non-hyperelliptic curves in M-theory. After the non-normalizable deformations, the gauge groups are decoupled and the
geometry has only conifold singularities. Now in the strong coupling regime, each gauge group has a gluino
condensation. Via a geometric transition which is obtained by shrinking $\P^1$ cycles and  making a normalizable
deformation,  the $\P^1$ cycles vanish and are replaced by the $\S^3$ cycles with RR, NS fluxes. The gluino
condensations are mapped into the sizes of  $\S^3$ cycles arising from  the normalizable deformation. The NS
fluxes, which are related to the couplings of the remaining Abelian gauge fields after gluino condensations, come
from the K\"ahler structure changes of the deformed geometry after the geometric transition while the RR fluxes are
originated from the vanished D5 branes.

By taking T-duality along the $U(1)$ direction of a natural $\C^*$ action on the $\CN=1$ geometry
(\ref{n1geometry}) given by \bea \lambda \cdot (x, y, u,v) \to (\lambda x, \lambda^{-1} y, u,v)~~\mbox{for $
\lambda \in \C^*$},\eea   one can obtain type IIA pictures which have been developed in a series of papers
\cite{dot1,dot2,dot3}. This T-duality gives the dictionary between the geometric engineering construction and the
Hanany-Witten type brane construction. D5 branes on the $\P^1$ cycles become D4 branes on intervals as $U(1)$ acts
along the angular direction of the $\P^1$'s and NS branes appear when the $U(1)$ orbits degenerate. When the NS
branes are projected onto the $u-v$ plane, they will be a collection of the holomorphic curves given by \bea u
\prod_{p=1}^n \left(u - \sum_{i=1}^p W'_i(v)\right) =0, \eea and their intersection points are in one-to-one
correspondence with the Higgsed gauge groups.

By lifting to eleven dimensional M theory, the brane configuration of D4 branes and NS branes in type IIA  becomes
a single M5 brane represented by a Riemann surface $\Sigma$ in a complex 3
dimensional space.
As the transition occurs in the limit where the sizes of $\P^1$ are very
small,
the information pertaining to the D4 branes is lost and we obtain an
M-theory plane M-theory curve which describes the remaining Abelian theory
after confinement. We then  go down to 10 dimensions and we get a brane configurations
with NS branes which can then be mapped to a deformed geometry with $\S^3$ cycles and fluxes after a T-duality.

Another interesting issue is on how to derive the remaining abelian gauge
theories which remain after confinement from the $\CN=2$ Seiberg-Witten
curves. The
point is that it is unknown  how to reduce the Seiberg-Witten curve
from
$\CN=2$ to $\CN=1$ in the presence of
bifundamental fields.  Nevertheless,
 we show how to obtain the $\CN=1$ parts and identify with the
M-theory curve after the transition.

Finally, we discuss orientifold theories which are obtained by complex conjugation on the geometry, Seiberg
dualities by introducing matter fields and birational flops and the relation to $G_2$ holonomy manifolds via
mirror symmetry.

\section{Vafa's Large N dualities}
\pagestyle{myheadings}
\markboth{Duality and Confiement in $N=1$...}{K. Oh and R. Tatar}

We will briefly review some of features of  Vafa's $\CN =1$ large N dualities which will be used later. Consider
type IIB theory on a non-compact Calabi-Yau threefold $O(-1) + O(-1)$ of $\P^1$ which is a small resolution of the
conifold : \bea xy -uv =0 \eea by wrapping $N$ D5 branes on $\P^1$. This gives a four dimensional $\CN=1$ $U(N)$
pure Yang-Mills theory described by open strings ending on the D5 branes, in the small 't Hooft parameter regime.
Vafa's duality states that in the large $N$ limit (large 't Hooft parameter regime), this is equivalent to type
IIB on the deformed conifold: \bea f= xy -uv -\mu =0. \eea In the deformed conifold, the $\P^1$ cycle is shrunken
and replaced by $\S^3$ of size $\mu$ which is identified with the condensate of the $SU(N)$ glueball superfield $S
= -\frac{1}{32 \pi^2} \mbox{Tr} W_\alpha W^\alpha$. The description is now in terms of closed strings. Rather than
the $N$ original D5 branes, there are now $N$ units of RR flux through $\S^3$, and also some NS flux through the
non-compact cycle dual to $\S^3$. The glueball $S$ is identified with the flux of the holomorphic 3-form on the
compact 3-cycle of the deformed conifold \bea S = \int_A \Omega \eea and the integral of  holomorphic 3-form on
the noncompact 3-cycle is made with introducing a cut-off $\La_0$: \bea \Pi = \int_B^{\La_0} \Omega =
\frac{1}{2\pi i}(-3S \log \La_0 -S + S \log S)\eea The effective superpotential is written as \bea W_{\mbox{eff}}
= \int (H_{RR} + \tau H_{NS}) \wedge \Omega \eea where $H_{RR}$ is the RR flux on the A-cycle and is due to the
$N$ D5 branes and  $H_{NS}$ is the NS flux on the noncompact B-cycle. By using the usual IR/UV identification in
the AdS/CFT conjecture, we identify the large distance  $\La_0$ (small IR scale) in supergravity with the small
distance (large UV scale) in the field theory such that the coupling constant in field theory is constant and
finite in UV. After doing so, the form of the effective superpotential is \bea W_{\mbox{eff}} = S \log
[\La^{3N}/S^N] +NS \eea The condition of supersymmetry implies that the derivative of $ W_{\mbox{eff}}$ with
respect to $S$ is zero which implies that $S$ gets $N$ discrete values, separated by a phase. This is the gluino
condensation in the field theory and signals the breaking of the chiral symmetry ${\bf Z_{2 N} \rightarrow Z_2}$.
The gluons of $SU(N)$ get a mass so the $SU(N)$ gets a mass gap and confines. What remains is the $U(1)$ part of
$U(N)$ whose coupling constant is equal to the coupling constant of the $U(N)$ theory divided by $N$.

We consider the lift of the transition to M theory
by using MQCD \cite{dot1,dot2,dot3}. A
T-duality of the geometrical constructions
takes the N D5 branes wrapped on $\P^1$ to a brane configuration
with two orthogonal NS branes on the directions $x$ and $y$ (together
with four directions corresponding to the
Minkowski space) and $N$ D4 branes in the the direction $x_n$.
The lift to M theory involves a single M5 brane
which has the worldvolume $R^{1,3} \times \Sigma$ where $\Sigma$ is
a 2-dim. manifold holomorphically embedded in
$(x, y, t)$ where $t = exp(\frac{x_n}{R_{10}} + i~x_{10})$.
When the $\P^1$ cycle shrinks, the direction $x_n$
goes to zero and eventually the coordinate $t$ becomes the coordinate
of a circle. Because we cannot embed
holomorphically into a circle, it results that the coordinate $t$ of the M5 brane become constant and $\Sigma$ is
embedded inside $x~y~=~\mbox{const.} $ where the constant is related to the scale of the $U(N)$ theory. After
reducing to ten dimensions,  $x~y~=~\mbox{const.} $ becomes the equations for a 2-dimensional surface where an NS
brane is wrapped, which is T-dual to the deformed conifold and the constant is related to the size of the $\S^3$
cycle. Therefore we could explicitly see the relation between the scale of the $U(N)$ theory and the size of the
$\S^3$ cycle.

The transition has been generalized to more complicated geometries in \cite{sv,civ,eot,ckv,GUD}, where the
blown-up geometry involves more
 $\P^1$ cycles and the deformed geometry involves more  $\S^3$ cycles. The
effective superpotential involves integrals of the holomorphic 3-form on the several  $\S^3$ cycles and several
noncompact cycles. Because the UV field theory involves several $U(N_i)$ groups, the running of  different gauge
group couplings gives rise to different bare couplings at the cut-offs on the noncompact cycles which corresponds
to different UV scales for the $U(N_i)$ groups. For different $U(N_i)$ groups in the $\CN =2$ theory we have
different $\be$-functions and the $\CN =2$ exact $\be$-function for the coupling $\tau _i\equiv {\theta _i\over
2\pi}+4\pi i g_i^{-2}$ of $U(N_i)$ is \bea \beta _i\equiv -2\pi i \beta (\tau _i)=\sum _j C_{ij}N_j,\eea with
$C_{ij}=2\delta _{ij}-|s_{ij}|=\vec e_i\cdot \vec e_j$ the Cartan matrix of the A-D-E diagram.

\section{$\CN =2$  A-D-E Quiver Gauge Theories}
\pagestyle{myheadings}
\markboth{Duality and Confiement in $N=1$...}{K. Oh and R. Tatar}

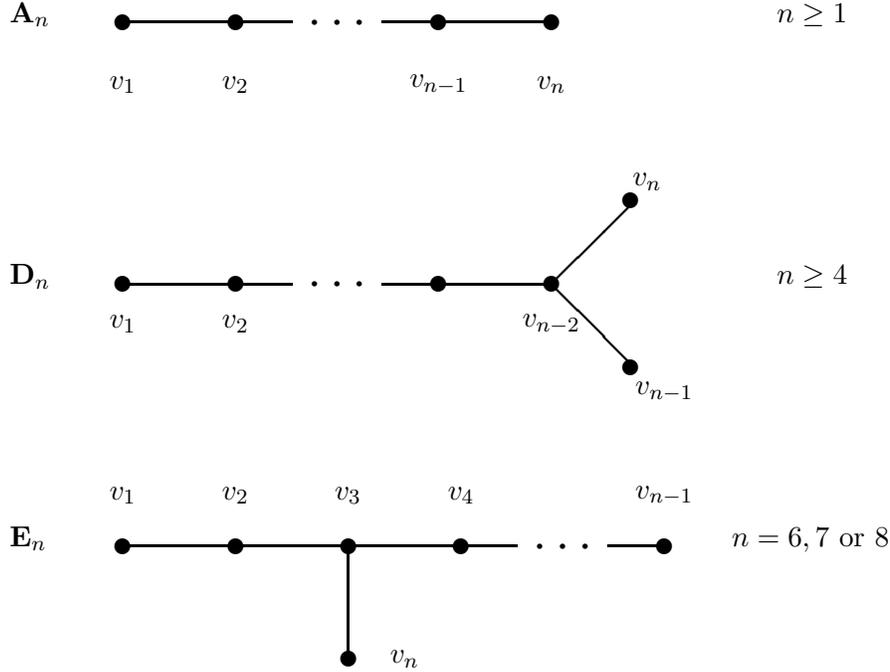
\begin{figure}
\setlength{\unitlength}{30mm}
\begin{center}

\begin{picture}(3.4,.9)(.5,.2)
\put(.5,1){$\A_n$}\thicklines \put(1,1){\line(1,0){.5}} \put(1,1){\circle*{.075}} \put(1.5,1){\circle*{.075}}
\put(1.5,1){\line(1,0){.25}} \put(1.85,1){\circle*{.02}} \put(1.95,1){\circle*{.02}} \put(2.05,1){\circle*{.02}}
\put(2.15,1){\line(1,0){.25}} \put(2.4,1){\circle*{.075}} \put(2.4,1){\line(1,0){.5}}
\put(2.9,1){\circle*{.075}}\put(.875,.6){\makebox(.25,.25){$v _ 1$}} \put(1.375,.6){\makebox(.25,.25){$v _ 2$}}
\put(2.275,.6){\makebox(.25,.25){$v _ {n-1}$}}\put(2.775,.6){\makebox(.25,.25){$v _ {n}$}}\put(3.9,1){$n\geq 1$}
\end{picture}

\begin{picture}(3.4,.9)(.5, .45)
\put(.5,1){$\D_n$}\thicklines \put(1,1){\line(1,0){.5}} \put(1,1){\circle*{.075}} \put(1.5,1){\circle*{.075}}
\put(1.5,1){\line(1,0){.25}} \put(1.85,1){\circle*{.02}} \put(1.95,1){\circle*{.02}} \put(2.05,1){\circle*{.02}}
\put(2.15,1){\line(1,0){.25}} \put(2.4,1){\circle*{.075}} \put(2.4,1){\line(1,0){.5}} \put(2.9,1){\circle*{.075}}
\put(2.9,1){\line(1,1){.35}} \put(3.25,1.37){\circle*{.075}} \put(2.9,1){\line(1,-1){.35}}
\put(3.25,0.63){\circle*{.075}} \put(.875,.7){\makebox(.25,.25){$v _ 1$}} \put(1.375,.7){\makebox(.25,.25){$v _
2$}} \put(2.775,.7){\makebox(.25,.25){$v _ {n-2}$}} \put(3.275,.4){\makebox(.25,.25){$v _ {n-1}$}}
\put(3.2,1.33){\makebox(.25,.25){$v _ n$}}\put(3.9,1){$n\geq 4$}
\end{picture}
\begin{picture}(3.4,.9)(1.4,.7)
\put(1.4,1){$\E_n$}\thicklines \put(1.9,1){\circle*{.075}} \put(1.9,1){\line(1,0){.5}} \put(2.4,1){\circle*{.075}}
\put(2.4,1){\line(1,0){.5}} \put(2.9,1){\circle*{.075}} \put(2.9,1){\line(0,-1){.5}} \put(2.9,.5){\circle*{.075}}
\put(2.9,1){\line(1,0){.5}} \put(3.4,1){\circle*{.075}} \put(3.4,1){\line(1,0){.25}} \put(3.75,1){\circle*{.02}}
\put(3.85,1){\circle*{.02}} \put(3.95,1){\circle*{.02}} \put(4.05,1){\line(1,0){.25}} \put(4.3,1){\circle*{.075}}
\put(1.775,1.1){\makebox(.25,.25){$v _ 1$}} \put(2.275,1.1){\makebox(.25,.25){$v _ 2$}}
\put(2.775,1.1){\makebox(.25,.25){$v _ 3$}} \put(3.025,.375){\makebox(.25,.25){$v_n$}}
\put(3.275,1.1){\makebox(.25,.25){$v _ 4$}} \put(4.175,1.1){\makebox(.25,.25){$v _ {n-1}$}} \put(4.6,1){$n=6,7$ or
$8$}
\end{picture}

\end{center}
\caption{A-D-E Dynkin Diagrams}\label{ADE}
\end{figure}

One can associate $\CN=2$ supersymmetric gauge theories with gauge group $\prod_{i=1}^n U(N_i)$ to each Dynkin
diagram of the simple complex Lie algebras of type $\A_n,\D_n, \E_n$ (Figure \ref{ADE}). Each factor $U(N_i)$
corresponds to a vertex $v_i$ in the Dynkin diagram and bifundamental hypermultiplet $Q_{i,j}$ correspond to  an
edge from $v_i$ to $v_j$. The bifundamental $Q_{i,j}$  is in the $(\N_i,\bar{ \N_j})$ representation of $U(N_i)
\times U(N_j)$.

The A-D-E Dynkin diagram also arise as a resolution graph of two dimensional quotient singularity $(\C^2/G, 0)$ by
a finite subgroup $G \subset SU(2)$, which is called A-D-E singularity.  The A-D-E singularity can be blown-up to
a smooth ALE space where the singular point is replaced by a configuration of rational curves $\P^1$. The
configuration can be explained in terms of the resolution graph  which consists of $n$ vertices corresponding to
the rational curves $\P^1$ and edges  between the vertices when the corresponding $\P^1$'s intersect. The
intersection matrix of the resolution graph is the negative of the Cartan matrix. The A-D-E singularities can be
embedded as hypersufaces $f(x,y,u)=0$ in ${\bf C}^3$: \bea \A_n:&\,\, &f=xy+u^{n+1}\cr \D_n:&\,\,
&f=x^2+y^2u+u^{n-1}\cr \E_6:&\,\,&f=x^2+y^4+u^3\cr \E_7:&\,\, &f=x^2+uy^3+u^3\cr
\E_8:&\,\,&f=x^2+y^5+u^3\label{ADEsingularities}\eea Furthermore, we can realize the $\CN =2$ A-D-E quiver gauge
theories on these resolved ALE spaces. Consider type IIB string theory compactified on the product of the resolved
ALE space and the flat complex plane $\C^1$. This product space can be view as the normal bundle $\CN$ of the
exceptional locus whose restriction to each $\P^1$ is $\CO(-2) \oplus \CO$ of $\P^1$. If we wrap $N_i$ D5 branes
around each $\P^1$, then we obtain the corresponding $\CN =2$ quiver gauge theories on the uncompatified
world-volume of the D5 brane which have been studied~\cite{kapustin}. The sections of the $\CO$ factor of
$\CN$ on each $\P^1$, which are given by the eigenvalues of the adjoints $\Phi_i$ of $U(N_i)$  are identified with
the Coulomb branch of the moduli.

\begin{figure}
\centerline{\epsfxsize=80mm\epsfbox{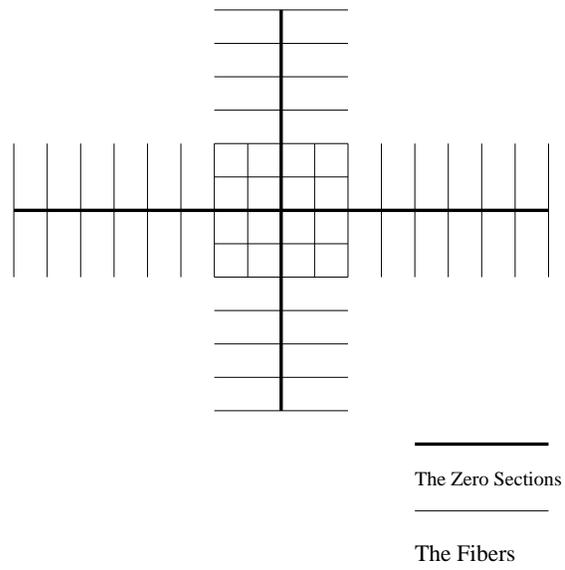}} \vspace*{1cm} \caption{The fibers of $\CO(-2)$ over $\P^1_i$
are identified with the sections of $\CO(-2)$ over $\P^1_{i+1}$ and vice versa.} \label{plumbing}
\end{figure}

We will now explicitly construct the resolved ALE space for the A-D-E singularities and the brane configuration
description via T-duality. The  resolutions of the A-D-E singularities can be  obtained by `plumbing' $n$
$\CO(-2)$ over $\P^1$. To be precise, we introduce two $\C^2$ for each $\CO(-2)$ of $\P^1$, denoted by
$\C_{i,0}^2$ and $\C_{i,\infty}^2$,
 whose coordinates are $ Z_i,\,\, Y_i$ for $\C_{i,0}^2$ and $ Z'_i,\,\, Y'_i $
for $\C_{i,\infty}^2$. Then the total space of the normal bundle $\CO(-2)$ over the $i$-th  $\P^1_i$ is given by
gluing $\C_{i,0}^2$ and $\C_{i,\infty}^2$ with the identification: \bea \label{o-2} Z'_i =1/Z_i, ~~Y'_i =Y_iZ_i^2,
\eea and $\P^1_i$ sits in as a zero section of the bundle. We denote the total space of $\CO(-2)$ of $\P^1_i$ by
$\mbox{Tot}(\CO(-2))$ and choose a real four dimensional
 small tubular neighborhood $\CT_i$ in  $\mbox{Tot}(\CO(-2))$ of the zero section $\P^1_i$
i.e. \bea \CT_i = \{ p\in \mbox{Tot}(\CO(-2)) | \mbox{dist}(p, (Z_i(p), Y_i(p) =0)) <\epsilon \}, \quad \mbox{for
a small }\, \epsilon \eea where $(Z_i(p), Y_i(p)=0)$ is the point on the zero section $\P^1_i$ obtained by the
projection along the fiber in $\mbox{Tot}(\CO(-2))$. We glue together $\CT_1, \cdots , \CT_n$ by plumbing a
neighborhood of the north pole ($Z'_i=X'_i=Y'_i=0$) of $\P^1_i$ in $\CT_i$ and a neighborhood of the south pole
($Z_{i+1}=X_{i+1}=Y_{i+1} =0$) of $\P^1_{i+1}$ in $\CT_{i+1}$ by exchanging the fiber (resp. base)
 coordinate $Y'_i$ (resp. $Z'_i$) of $\CO(-2)$ over the $i$-th $\P^1$ with the base (resp. fiber)
coordinate $Z_{i+1}$ (resp. $Y_{i+1}$) of $\CO(-2)$ over the $(i+1)$-th $\P^1$ (Figure \ref{plumbing}).
\\
\\
 In other
words, the plumbing is an isomorphism between portions of $\CT_i$ and $\CT_{i+1}$
induced by the map \bea
Y'_i \to Z_{i+1},\,\,\, Z'_i \to Y_{i+1}.\eea Note that the north pole  of the $i$-th $\P^1$ will
meet the south pole of the $(i+1)$-th $\P^1$ after the plumbing. Then
the minimal resolution of the $\A_n$ singularity is
isomorphic to a union of the tubular neighborhoods \bea \label{anresol} \CT = \CT_1 \cup \ldots \cup \CT_n \eea
with the plumbing. For other A-D-E resolutions, we glue $\CT_i$ following the Dynkin diagram.

\begin{figure}
\centerline{\epsfxsize=100mm\epsfbox{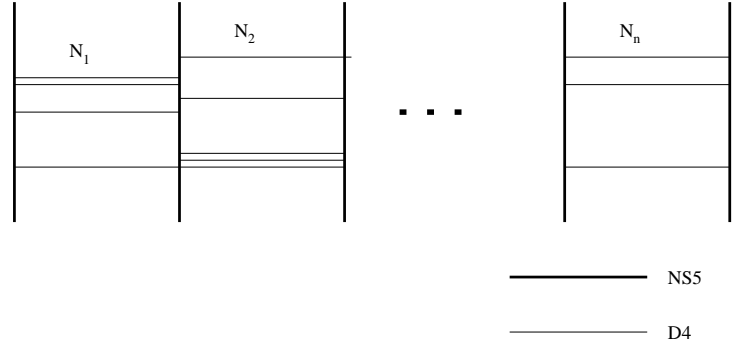}} \vspace*{1cm} \caption{$\A_n$ brane configuration: D5 branes
wrapping $\P^1$ cycles are T-dualized to D4 branes between NS branes.} \label{multi}
\end{figure}

Consider  a  circle  action  $S_2$  on  $\CT_i $: \bea \label{anaction}(e^{i\theta}, Z_i) = e^{i\theta}Z_i,
~~(e^{i\theta}, Y_i) = e^{-i\theta}Y_i\\ \nonumber (e^{i\theta}, Z'_i) = e^{-i\theta}Z'_i, ~~(e^{i\theta}, Y'_i) =
e^{i\theta} Y'_i.\eea Then this action is compatible with the plumbing because the plumbing exchanges $Z'_i$ with
$Y_{i+1}$ and $Y_i$ with $Z_{i+1}$, and thus globalizes to the action on the $\CT$. Since the orbits of the action
degenerate along $Z_i=Y_i=0$ and $Z'_i=Y'_i=0$, we have two NS branes along the transversal direction to $\CT$ at
$Z_i=Y_i=0$ and $Z'=Y'=0$ on each open set $\CT_i$ after T duality. We will then have $(n+1)$ NS branes labelled
from 0 to $n$, which are parallel because the transversal direction to $\CT$ is chosen to be flat.  Thus the
T-dual of $N_i$ D5 branes wrapping $\P^1_i$ of the resolution of $\A_n$ singularity will be a brane configuration
of $N_i$ D4 branes between the $(i-1)$-th and $i$-th NS branes as in Figure \ref{multi}. The method works for
other A-D-E singularities.
\\

\section{Field Theory Analysis-Classical Vacua}
\pagestyle{myheadings}
\markboth{Duality and Confiement in $N=1$...}{K. Oh and R. Tatar}

We now consider deformations of the $\CN =2$ gauge quiver theories to $\CN =1$ supersymmetric theory by adding  a
tree-level superpotential, which is in general of the form \bea \label{supotg} W =\sum_{i=1}^{n} W_i
-\mbox{Tr}\sum_{i=1}^{n}\sum_{j=1}^n s_{i,j}Q_{i,j} \Phi_{j} {Q}_{j,i},~~\mbox{where }W_i =
\mbox{Tr}\sum_{j=1}^{d_i+1} \frac{g_{i,j-1}}{j} \Phi_i^j,
\end{eqnarray}
and $s_{i,i} =0$ and $s_{i,j} = -s_{j,i} =1 $ if there is an edge between $v_i, v_j$ in the Dynkin diagram which
means that the corresponding $\P^1$ cycles intersect.~\footnote{The choice of the sign for $s_{i,j}$ does not make
any differences and we will assume $s_{i,j} =1$ if $i<j$ in the case of $\A_n$ or $\D_n$.} The matrix  $C_{ij} = 2
\delta_{ij} - |s_{ij}|$ is the Cartan matrix of the A-D-E diagram. As pointed out in \cite{GUD, GK}, for $m>1$ the
non-renormalizable interaction terms seem to be irrelevant for the long distance behavior of the theory, but these
terms have in general strong effects on the infrared dynamics. These are examples of operators known as
`dangerously irrelevant'.

In this section, we find the supersymmetric classical vacua by solving D-term and F-term equations. The D-flat
condition implies that:
 \bea
 \label{Dterm1}
D^a_i =\mbox{Tr}\left( \Phi_i^\dagger t_a^{(\bf{N_i})}\Phi_i + \Phi_i t_a^{(\bf{\bar{N}_i})}\Phi_i^\dagger +
\sum_j s_{ij}\left( Q_{i,j}^\dagger t_a^{(\bf{N_i})}Q_{i,j} - Q_{j,i}^\dagger t_a^{(\bf{\bar{N}_i})}Q_{j,i}
\right)\right) =0\eea where $t_a^{(\bf{N_i})}$ are the generators in the fundamental of $U(N_i)$ and
$t_a^{(\bf{\bar{N}_i})}$ are those in the anti-fundamental of $U(N_i)$. This can be rewritten as: \bea
\label{Dterm} &[\Phi_i, \Phi_i^\dagger] = 0,\\\label{Dtermsecond} & s_{ij} \left(Q_{i,j}Q_{i,j}^\dagger -
Q_{j,i}^\dagger Q_{j,i} \right) = -\eta_i \mbox{Id}_{N_i}.\eea  As noticed in \cite{Luty:1995sd, 9603042}, the
reason for this is that imposing D-terms equations together with the gauge equivalences is the same as taking the
quotient under the complexified gauge groups which can used to diagonalize $\Phi_i$, thus automatically
satisfying (\ref{Dterm}). In this context, the D-terms equations are the moment map in symplectic quotient.
(\ref{Dtermsecond}) can be proved inductively on $i$ for the A-D-E quiver theory. In the case of the $\A_n$
theory, as shown in the appendix of \cite{csaki} that, after color rotations, the $Q_{i,j}$ can be simultaneously
diagonalized, so that they are of the form \bea Q_{i,j} = \mbox{diag}\{q^{(a)}_{i,j}\}, \quad Q_{j, i}
=\mbox{diag} \{{q}^{(a)}_{j,i}\} \eea with \bea |{q}^{(a)}_{j,i}| = |q^{(a)}_{i,j}|, \quad a =1,\ldots ,
N_{i,i+1} := \mbox{min }(N_i, N_{i+1}). \eea  For the $\D_n$ or $\E_n$ theories, this does not hold in
general.~\footnote{We thank the referee for pointing out this.} In the following discussions, the diagonalization
of $Q_{i,j}$ is not used in an essential way. Sometimes we will write $q^{(a)}_i$ for $q^{(a)}_{i,j}$ and
$\tilde{q}^{(a)}_i$ for $q^{(a)}_{j,i}$ when $j=i+1$ if there is no confusion.

We now look at  the F-term equations $d W=0$. The equations are
\bea\label{Ftermg} &s_{i,j}(Q_{i,j}
\Phi_{j} - \Phi_{i} Q_{i,j}) =0, \\ \non&W'_i (\Phi_i) = \sum_j s_{ij}Q_{i,j} Q_{j, i}
\eea The first equation also implies that \bea s_{i,j} ( Q_{i,j}Q_{j,i} \Phi_i -\Phi_i Q_{i,j}Q_{j,i} )= 0,\eea
From this and the second  equation, it is clear that the unbroken gauge symmetry of  $\Phi_i$ will also preserve
$Q_{i,j} Q_{j,i}$. Therefore, we can solve the F-term equations in terms of the unbroken gauge symmetries of
$Q_{i,j}Q_{j,i}$, which are determined by the eigenvalues of $Q_{i,j}Q_{j,i}$. For each factor group $U(N_i)$, we
restrict the F-term equations on the gauge invariant subspace for which the eigenvalues of the mesons
$Q_{i,j}Q_{j,i}$ is fixed and we can divide into two cases.~\footnote{We will restrict to the $\A_n$ quiver
theory, the method works also for the $\D_n$ and $\E_n$ quiver theories.}

$\bullet$  {\bf Case 1.} On the gauge invariant subspace for which the eigenvalues of  $Q_{i,j}Q_{j,i}$ are zero
for all $j$.

Let $U(M_{i,i})$ be the maximal gauge symmetry subgroup of $U(N_i)$ for which the eigenvalues of $Q_{i,j}$ are
zero. Then (\ref{Ftermg}) implies that \bea W'_i (\Phi_i) =0, \eea on the subspace invariant under $U(M_{i,i})$.
Hence the eigenvalues $v_i$ of $\Phi_i$ is given by \bea \label{phieigen} W'_i(v_i) = \sum_{l=1}^{d_i} g_{i,l}
v_i^l :=g_{i,d_i} \prod_{l=1}^{d_i} (v_i - b_{i,l}) =0,\eea and the gauge group will break \bea U(M_{i,i})
\rightarrow \prod_{l=1}^{d_i}U(M_{i,i,l}).\eea

$\bullet$ {\bf Case 2.} On the gauge invariant subspace for which the eigenvalues of $Q_{i,j}Q_{j,i}$  are
non-zero for some $j$. Then for a given non-zero eigenvalue of $Q_{i,j}Q_{j,i}$ , there exist $j\leq i < k$ such
that \bea\prod_{l=j}^{k-1}Q_{l,l+1}\prod_{l=1}^{k-j}Q_{k-l+1,k-l}\eea have
non-zero eigenvalues and \bea\prod_{l=j-1}^{k-1}Q_{l,l+1}\prod_{l=1}^{k-j+1}Q_{k-l+1,k-l}=0,\\
\prod_{l=j}^{k}Q_{l,l+1}\prod_{l=0}^{k-j}Q_{k-l+1,k-l}=0.\eea Let $U(M_{j,k})$ be the maximal gauge symmetry
subgroup for the given non-zero eigenvalue of $Q_{i,j}Q_{j,i}$. We first note that $U(M_{j,k})$ embeds diagonally
into the factor $\prod_{i=j}^k U(N_i)$ of $ \prod_{i=1}^n U(N_i)$ as all bifundamentals $Q_{l,l+1}, Q_{l+1,i}$ for
$l=j, \ldots, k-1$ have the unique non-zero expectation value. The first equation of (\ref{Ftermg}) implies that
the expectation values of $\Phi_m$ for $m=j, \ldots, k$ are the same, and hence from the second F-term equation
(\ref{Ftermg}), we see that the eigenvalue values $v_{j,k}$ of $\Phi_m$ must satisfy \bea\label{wjk} \sum_{l=j}^k
W_l'(v_{j,k}) = 0. \eea Generically, the number of different solutions is given by the maximum degree of $W_l'$:
\bea \label{djk} d_{j,k} =\max \{ d_j, d_{j+1}, \ldots, d_k \}. \eea By considering the solutions
$v_{j,k,l},~~l=1,\ldots , d_{j,k}$ of (\ref{wjk}), the expectation values $q_{m,m+1,l}{q}_{m+1,m,l}$ of the mesons
$Q_{m,m+1}Q_{m+1,m}$ (i.e. of the bifundamentals) are determined uniquely from the second F-term equation
(\ref{Ftermg}):\bea\label{bifundexp} q_{m,m+1,l}{q}_{m+1,m,l} = \sum_{i=j}^{m} W_i'(v_{j,k,l}).\eea Then the gauge
group is generically Higgsed to \bea U(M_{j,k}) \rightarrow \prod_{l=1}^{d_{j,k}}U(M_{j,k, l}), \eea and
$U(M_{j,k,l})$'s are diagonally embedded into $U(N_j) \times U(N_{j+1}) \times \cdots U(N_k)$.  There are
$n(n-1)/2$ possibilities for $U(M_{j,k})$. Combining with {\bf Case 1}, the number of gauge group breakings is
\bea \label{d} d:= \sum_{j=1}^n \sum_{k=j}^n d_{j,k} \eea where $d_{ii} = d_i$. If all degrees of the adjoint
superpotential term $W_i$ are equal to the same value $p$, then there are $p(n+1)n/2$ breakings which agrees with
\cite{ckv}. Therefore in the case of $\A_n$ quiver theory, there are finitely many discrete branches of the vacua
parameterized by \bea M_{j,k,l}, ~~j=1,\ldots, n,~~k=j, \ldots, n,~~l=1,\ldots , d_{j,k}\eea for which the gauge
group is Higgsed to: \bea \prod_{i=1}^n U(N_i) \rightarrow \prod_{j=1}^n\prod_{k=j}^n\prod_{l=1}^{d_{j,k}}
U(M_{j,k,l}), \eea where $d_{j,k}$ are defined in (\ref{djk}), provided that a set of  equations \bea
\label{genericcond} W_j'(v) + W'_{j+1}(v) +\ldots + W_k'(v) =0 \eea have exactly $d_{j,k}$ different solutions.
This particularly means that any two curves \bea u- \sum_{i=1}^{j-1} W_i'(v) =0,~~  u - \sum_{i=1}^k W_k'(v) =0
\eea meet transversally at exactly $d_{j,k}$ different points. The equation $(\ref{genericcond})$ is equivalent to
the condition that the system \bea\non
\frac{d^2 \sum_{i=j}^k  W_i(v)}{dv^2} =0,\\
\frac{d \sum_{i=j}^k  W_i(v)}{dv} =0. \eea does not have any common solution, and $\sum_{i=j}^k  g_{i,d_i} \neq
0.$

 We will now consider two particular cases of the superpotential
(\ref{supotg}) in great details, which will be called
 quadratic  and degenerate superpotential, respectively:
\begin{eqnarray}
\label{supot} \Wq &= &\mbox{Tr}\left(\sum_{i=1}^{n} (\frac{g_i}{2} \Phi_i^2 + h_i \Phi_i )-
\sum_{i=1}^{n}\sum_{j=1}^n s_{i,j}Q_{i,j} \Phi_{j} {Q}_{j,i} \right)\\ \label{supotd} \Wd &=&
\mbox{Tr}\left(\sum_{i=1}^{n} \frac{g_i}{m+1} \Phi_i^{m+1} - \sum_{i=1}^{n}\sum_{j=1}^n s_{i,j}Q_{i,j} \Phi_{j}
{Q}_{j,i} \right)\
\end{eqnarray}
The $\CN=1$ theory deformed by (\ref{supotd}) or (\ref{supot}) contains
certain universal IR aspects of theories with more general superpotential
(\ref{supotg}), while they are amenable to a more concrete and detailed
analysis due to their simplicity and they will be also a basis for our
analysis of the general case. We assume
that $g_{n-1} =g_n$ for $\D_n$ theory which means that we are giving the
same mass to the last two adjoints
$\Phi_{n-1}, \Phi_n$. We do this in order to preserve the $\Z_2$ symmetry of exchanging
the last two vertices in
the $\D_n$ Dynkin diagram.

$\bullet$ The case of the $\A_n$ quiver theory with  quadratic  superpotential $\Wq$ (\ref{supot}):

We may rewrite the F-term equation (\ref{Ftermg}) as \bea \label{Ftermq} &Q_{i,i+1} \Phi_{i+1} - \Phi_{i}
Q_{i,i+1} =0,~~i=1, \ldots, n-1,
\\ \non& g_i\Phi_i +h_i  = Q_{i,i+1} Q_{i+1, i} -Q_{i,i-1}Q_{i-1,i},~~i=1, \ldots ,n
\eea where $Q_{1,0}, Q_{0,1}, Q_{n,n+1}, Q_{n+1,n}$ are defined to be zero.
By substituting the adjoints $\Phi_i$ in the first equation in terms of
the mesons as given by the second equation, we conclude that
either $ q^{(a)}_{i} =0$
or
 \bea \label{fterm}   {g_{i+1}}q^{(a)}_{i-1}\tilde{q}^{(a)}_{i-1} -
 \left( {g_{i+1}} +{g_{i}} \right) q^{(a)}_{i} \tilde{q}^{(a)}_{i}  + {g_{i}} q^{(a)}_{i+1}\tilde{q}^{(a)}_{i+1} =
{h_{i+1}}{g_{i}} - {h_i}{g_{i+1}} ,\eea for $i=1, \ldots, n$ and $a = 1, \ldots, N_{i,i+1}$.
Here the indexed
quantity is assumed to be zero if the lower index is not between 1 and $n$.
For the $\D_n$ theories, the equations (\ref{fterm}) for $i=n-1, n$
should be replaced by 
\bea &{g_{n-1}}q^{(a)}_{n-3}\tilde{q}^{(a)}_{n-3} - \left(
{g_{n-1}} + {g_{n-2}} \right) q^{(a)}_{n-2} \tilde{q}^{(a)}_{n-2}  -{g_{n-1}} q^{(a)}_{n-2,n}{q}^{(a)}_{n,n-2} \nonumber \\&\hspace{3in}=
h_{n-1}{g_{n-2}} - h_{n-2}{g_{n-1}},
\nonumber\\
& {g_{n}}q^{(a)}_{n-3}\tilde{q}^{(a)}_{n-3} - \left( {g_{n}} + {g_{n-2}} \right) q^{(a)}_{n-2,n} {q}^{(a)}_{n,n-2}
-{g_{n}} q^{(a)}_{n-2}\tilde{q}^{(a)}_{n-2}  = {h_{n}}{g_{n-2}} - {h_{n-2}}{g_{n}}. \quad \eea
 When $q^{(a)}_j
q^{(a)}_{j+1} \cdots q^{(a)}_{k-1} \neq 0$ for $j< k$ and $q^{(a)}_{j-1} =
q^{(a)}_{k} =0$, the equations (\ref{fterm}) become a system of $(k-j)$
linear equations in $(k-j)$ unknowns $q^{(a)}_{j}\tilde{q}^{(a)}_{j},
\ldots , q^{(a)}_{k-1}\tilde{q}^{(a)}_{k-1}$: \bea \label{linsys} &\left(g_{j} + {g_{j+1}} \right) q^{(a)}_{j}
\tilde{q}^{(a)}_{j} - {g_{j}} q^{(a)}_{j+1}\tilde{q}^{(a)}_{j+1} = h_j{g_{j+1}}- h_{j+1}{g_{j}}.
\\ \nonumber &- {g_{j+2}}q^{(a)}_{j}\tilde{q}^{(a)}_{j} + \left(
{g_{j+1}} +
{g_{j+2}} \right) q^{(a)}_{j+1} \tilde{q}^{(a)}_{j+1} - {g_{j+1}} q^{(a)}_{j+2}\tilde{q}^{(a)}_{j+2} =  h_{j+1}{g_{j+2}} -h_{j+2}{g_{j+1}}. \\
\nonumber& \vdots\qquad\qquad\qquad
 \qquad\qquad\qquad\qquad \\
\nonumber  &- {g_{k}}q^{(a)}_{k-2}\tilde{q}^{(a)}_{k-2} + \left( {g_{k-1}} + {g_{k}} \right) q^{(a)}_{k-1}
\tilde{q}^{(a)}_{k-1}  - {g_{k-1}} q^{(a)}_{k}\tilde{q}^{(a)}_{k} =h_{k-1}{g_{k}} - h_{k}{g_{k-1}}. \eea
Hence it has a unique solution if the system is linearly independent
which is equivalent to having a non-zero determinant of the following
$(k-j+1)\times (k-j+1)$ matrix: \bea \label{determinant}\left|
\begin{array}{ccccccc}
-g_{j+1} & g_j & 0 &0& \ldots & 0&0\\
0& -g_{j+2} & g_{j+1} & 0 & \ldots &0&0\\
\vdots& \vdots &\vdots & \vdots  &\vdots &\vdots &\vdots \\
0& 0& 0&0& \ldots & -g_{k}& g_{k-1} \\
1& 1& 1&1&\ldots &1&1
\end{array}
\right|. \eea Here only two entries in each row are non-zero except in the last row
where all entries are $1$'s.
Thus either $q_i$'s are zero or if $q^{(a)}_j q^{(a)}_{j+1} \cdots q^{(a)}_{k-1} \neq 0$ for $j< k$ and
$q^{(a)}_{j-1} = q^{(a)}_{k} =0$, there is a unique non-zero solution. As
the components of $Q_{j,j+1}, \ldots, Q_{k-1, k}$ have non-zero
expectation values, the gauge group is Higgsed to \bea \prod_{i=1}^n U(N_i) \rightarrow
\prod_{i=1}^{j-1} U(N_i) \times U(M_{j,k})\times\prod_{i=j}^k U(N_i -M_{j,k})\times \prod_{i=k+1}^n U(N_{i}), \eea
where $M_{j,k} \leq \mbox{min}\,\{N_i, N_{i+1}, \ldots, N_{k}\}$. The
subgroup $U(M_{j,k})$ is diagonally embedded in $U(N_j) \times \cdots
\times U(N_k)$ such that when $Q_{l,l+1}Q_{l+1,l}$ is restricted to the
fundamental and anti-fundamental representations of $U(M_{j,k})$ factor of  $U(N_l)$,
it is a diagonal matrix with the same diagonal entries:
\bea
Q_{l.l+1}Q_{l+1,l} = \mbox{diag} \{{q}_l \tilde{q}_l\}, \quad\mbox{for } l =j
,\ldots ,k-1
\eea
where $q_l \tilde{q}_l$ is the solution of (\ref{linsys}).

$\bullet$ Moduli space of $\A_3$ quiver theory:

We first consider the $\A_3$ quiver theory. After possible
changes of bases, we may assume that \bea \label{q12} Q_{1,2}Q_{2,1}&=& \mbox{diag} \{ p^{(1)}_1, \ldots,
p^{(M_{1,3})}_1, q^{(1)}_1, \ldots , q^{(M_{1,2})}_1,\,
 0, \ldots, 0 \}\\
\label{q23} Q_{2,3}Q_{3,2}&=& \mbox{diag} \{ p^{(1)}_2, \ldots, p^{(M_{1,3})}_2, \underbrace{0, \ldots,
0}_{M_{1,2}},\, q^{(1)}_2, \ldots , q^{(M_{2,3})}_2,\, 0, \ldots, 0 \} \eea where \bea &p^{(1)}_1=  \ldots=
p^{(M_{1,3})}_1& = \frac{h_1g_2 -g_1h_2 +h_1g_3 -g_1h_3}{g_1+g_2 +g_3}
,\non\\
&q^{(1)}_1= \ldots = q^{(M_{1,2})}_1 &= \frac{g_2h_1 -g_1h_2}{g_1 + g_2},\non\\
&p^{(1)}_2 = \ldots = p^{(M_{1,3})}_2 &= \frac{h_2g_3 -g_2h_3 +h_1g_3 -g_1h_3}{g_1+g_2+g_3},\non\\
&q^{(1)}_2= \ldots = q^{(M_{2,3})}_2 &= \frac{g_3h_2-g_2h_3}{g_2 + g_3}. \eea This will imply that
 \bea
\label{q13} \Phi_1 & =& \frac{1}{g_1}(Q_{12}Q_{21} -h_1\mbox{Id})\non
\\\Phi_2 &=& \frac{1}{g_2}(Q_{23}Q_{32} - Q_{21}Q_{12} - h_2\mbox{Id})\non
\\\Phi_3 &=& \frac{1}{g_3}(-Q_{32}Q_{23} - h_3\mbox{Id})
\eea
With the above choice for the Higgsing, the gauge group is broken to:
\bea
\lefteqn{ U(N_1) \times U(N_2) \times U(N_3)}\\ \nonumber & &\rightarrow
U(N_1 -M_{1,2}-M_{1,3}) \times U(N_2 -M_{1,2}- M_{2,3} - M_{1,3}) \times U(N_3 -M_{2,3}-M_{1,3})\\
\nonumber & &\times U(M_{1,2}) \times U(M_{2,3}) \times U(M_{1,3}). \eea where $U(M_{1,2})$ is diagonally embedded
into the product $U(N_1) \times U(N_{2})$, $U(M_{2,3})$ into $U(N_2) \times U(N_3)$, and $U(M_{1,3})$ into $U(N_1)
\times U(N_2) \times U(N_3)$ respectively.

Generally, in the case of $\A_n$ quiver theory, there are finitely many discrete branches of the vacua
parameterized by $M_{j,k}, j=1,\ldots ,n, k =j+1, \ldots, n$ in which  the gauge group is broken as follows:
\bea  \label{Higgs} \prod_{i=1}^n U(N_i) \rightarrow \prod_{j=1}^{n}\prod_{k=j}^{n} U(M_{j,k}),\\
\,\,\mbox{where}\,\,\,M_{i,i} = N_i - \sum_{\stackrel{j\leq i \leq k}{j\neq k}}^{\,\,} M_{j,k}.\nonumber\eea Here
$U(M_{j,k})$ is the largest group which can be diagonally embedded into $U(N_j) \times U(N_{j+1}) \times \cdots
\times U(N_{k})$ of $\prod_{i=1}^n U(N_i)$, but cannot be embedded into a larger group $\prod_{i=1}^n U(N_i) $.
The  group $U(M_{j,k})$ appears when the blocks of $Q_{j,j+1}, \ldots , Q_{k-1,k}$ have simultaneously non-zero
expectation values. Therefore the product of $n$ gauge groups is broken into the product of $\frac{n(n+1)}{2}$
gauge groups in a generic branch in the $\A_n$ quiver theory. Notice that the vacua is completely determined by
the following set of the complex lines in the $u-v$ space up to discrete moduli $M_{j,k}$: \bea u - \sum_{j=1}^i
(g_i v +h_i)=0, \,\ , i =0, 1, \ldots, n, \eea where we set $g_0=h_0 =0$. Conversely, these lines are determined
by the vacua. There are only $2n$ free continuous parameters in this correspondence and the vacua cannot be
arbitrary, we will interpret these lines as NS branes and the intersection points of these lines as the location
of D-branes in type IIA picture in the next section (Figure \ref{A3NS}).

$\bullet$ The case of the $\D_n$ quiver theory with  quadratic  superpotential $\Wq$ (\ref{supot}):

In the $\D_n$ quiver theory studied in \cite{Gubser:1998ia}, the F-term equations are the same as in the $\A_n$
quiver theory except the last two, corresponding to the roots $v_{n-1}, v_n$ and there is no difference in the
procedure of finding the vacua. Because of our mass assumption $g_{n-1} = g_n$, we will have one less gauge group
breaking in the generic Higgs branch. So the vacua are parameterized by $n(n+1)/2 -1$ natural numbers. In the
branch parameterized by $M_{j,k}$, the gauge group is Higgsed to: \bea \prod_{i=1}^n U(N_i) \rightarrow
\prod_{j=1}^{n-1}\prod_{k=j}^{n-1}U(M_{jk}) \times \prod_{j=1}^{n-2}U(M_{j,n}),\\
\mbox{where}\,\,\, M_{i,i} = N_i - \sum_{\stackrel{j\leq i \leq k}{j\neq k}}^{\,\,} M_{j,k}.\nonumber\eea In terms
of the line configurations in the $u-v$ plane, one missing gauge group in the Higgsing is due to the fact that two
lines corresponding to the last two vertices of the $\D_n$ diagram are parallel.

$\bullet$ The case of the $\A_n$ quiver theory with degenerate superpotential $\Wd$ (\ref{supotd}):

The F-term equation (\ref{Ftermg}) can be rewritten as follows: \bea \label{Ftermd} & Q_{i,i+1}\Phi_{i+1} - \Phi_i Q_{i,i+1} =0,\\
\non &g_i \Phi_i^m = Q_{i,i+1}Q_{i+1,i} - Q_{i,i-1}Q_{i-1,i},~~i=1, \ldots , n \eea where $Q_{1,0}, Q_{0,1},
Q_{n,n+1}, Q_{n+1,n}$  are defined to be zero. The first equation also implies that \bea \label{Ftermqm}Q_{i,j}
\Phi_{j}^m- \Phi_{i}^m Q_{i,j}=0. \eea
So, by eliminating the $\Phi_i^m$, we conclude that the bifundamentals
$Q_{i,i+1}, Q_{i+1,i}$ are zero and hence the solutions are
\bea \label{solutiondeg} Q_{i,i+1}=0,~~ Q_{i+1,i} =0,
~~\Phi_i^m=0,\eea provided that the condition (\ref{determinant}) for $g_i$'s are satisfied. Classically this
means that $\Phi_i=0$, but at quantum level $\Phi_i$ is allowed to be a nilpotent i.e. $\Phi^m =0$.  The nilpotent
solutions $\Phi_i^k = 0, \Phi_i^{k-1} \neq 0, k=2, \ldots, m$ are the infinitesimal deviations from the vacuum at
$\Phi_i =0$, which means that the $\P^1$ cycles on which D5 branes wrap on can be deformed in an infinitesimal
neighborhood. In terms of the brane configuration, there will be $m$ NS branes infinitesimally deviated from each
other which bound the infinitesimally deviated D4 branes. We will revisit these issues later.

\section{$\CN =1$ theory - Brane Interpretation}
\pagestyle{myheadings}
\markboth{Duality and Confiement in $N=1$...}{K. Oh and R. Tatar}

We give a brane configuration interpretation of the results of the previous section  for the $\A_n$ quiver theory.
Before we begin our discussion with the arbitrary superpotential, we illustrate the idea for the case of the
$\A_2$ quiver theory with  the quadratic superpotential $\Wq$ (\ref{supot}). We will also consider the $\D_n$
theory  and deal with the degenerate superpotential in the next section.

\begin{figure}
 \centerline{\epsfxsize=120mm\epsfysize=70mm\epsffile{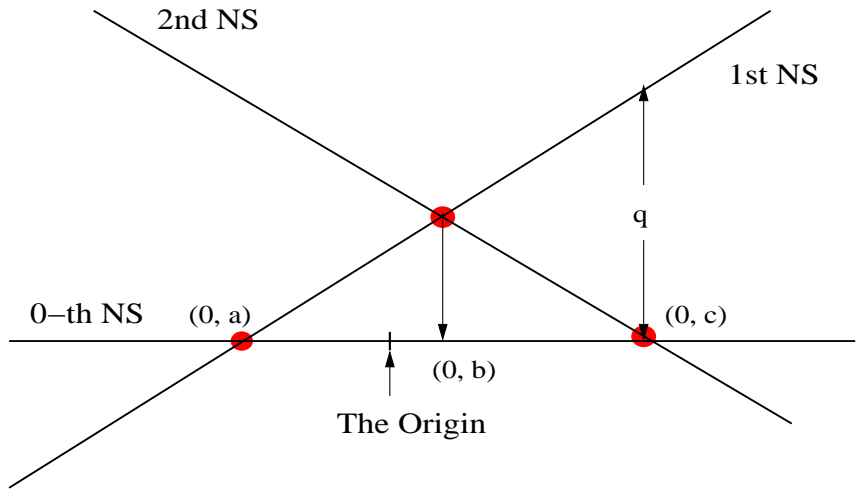}}
\caption{A brane configuration for the $\CN=1$ $\A_2$ theory with the quadratic superpotential. Here $a$ (resp.
$b$) is the expectation value of $\Phi_1$ (resp. $\Phi_2$) on $U(M_{1,1})$ (resp. $U(M_{2,2})$), $c$ is the common
expectation value of $\Phi_1$ and $\Phi_2$ on $U(M_{1,2})$ and $q$ is the expectation value of the meson
$Q_{1,2}Q_{2,1}$.  $\bullet$'s are the location of  D4 branes.}\label{A3NS}
\end{figure}

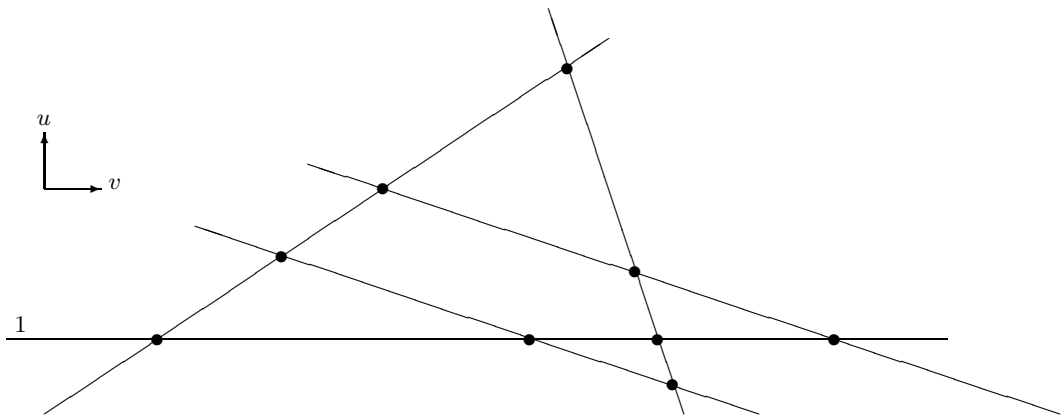
\begin{figure}
\setlength{\unitlength}{0.5mm}
\begin{picture}(500, 150)(0,0)
\put(20, 40){\line(1,0){250}} \put(30, 80){\vector(1,0){15}}\put(47,80){\footnotesize$v$}
\put(30,80){\vector(0,1){15}}\put(28, 97){\footnotesize$u$}\put(22,42){\footnotesize$1$} \put(30,
20){\line(3,2){150}} \put(200, 20){\line(-1,3){36}}\put(300, 20){\line(-3,1){200}} \put(220, 20){\line(-3,1){150}}
\put(157,38){$\bullet$}\put(195,26){$\bullet$}\put(91,60){$\bullet$}
\put(58,38){$\bullet$}\put(191,38){$\bullet$}\put(238, 38){$\bullet$}
\put(118,78){$\bullet$}\put(185,56){$\bullet$}\put(167,110){$\bullet$}
\end{picture}
\caption{The brane configuration in $u-v$ space for $\D_n$.  Note that the last two NS branes are
parallel.}\label{dnNS}
\end{figure}

Recall that the $\CN =2$ brane configuration, where D4 branes with three NS branes, denoted as zeroth, first and
second from the left to the right, can move freely along the direction of NS branes which was denoted as
$v$-direction. Therefore we can identify the NS branes with the moduli of the positions of the D4 branes. When the
superpotential (\ref{supot}) is introduced, the expectation values of $\Phi_i,~~i=1,2$ are determined by \bea W_i'
=g_i(v+\frac{h_i}{g_i})=0\eea when the expectation value of the meson $Q_{1,2}=Q_{2,1}$ is zero. We denoted the
maximal unbroken gauge symmetry groups by $U(M_{i,i})$ at the vacua with these expectation values. Hence the
$M_{ii}$ of $N_i$ D4 branes stretched between $(i-1)$-th and $i$-th NS branes will be fixed at the position $v=
-h_i/g_i$ in the vacua. When the expectation values of the meson $Q_{1,2}Q_{2,1}$ is non-zero, there is a common
expectation value of the adjoints $\Phi_1$ and $\Phi_2$ given by \bea W_1'(v) + W_2'(v) =(g_1 +g_2)(v + \frac{h_1
+h_{2}}{g_1+g_{2}})= 0. \eea The maximal unbroken gauge group is denoted by $U(M_{1,2})$ which is diagonally
embedded into $U(N_1) \times U(N_2)$. It means that the $M_{1,2}$ D4 branes connecting the 0-th and the first NS
and $M_{1,2}$ D4 branes from connecting the first and the second are merged together  at $v= - \frac{h_1
+h_{2}}{g_1+g_{2}}$ so that they form a long D4 brane stretched from the zeroth to the second NS branes.
 Also, by giving the masses to the adjoints $\Phi_i$, the NS branes
are rotated. It seems impossible to do so because D4 branes between adjacent NS branes are split into two stacks
rather than one stack so the supersymmetry would not be preserved. But if we make a translation of the first and
the second NS branes into the $u$ direction, one can rotate the NS branes. This translation in $u$ direction is
due to the non-zero expectation value of the meson $Q_{1,2}Q_{2,1}$. Thus we identify the $v$-direction with the
freedom of the adjoints moving along the Coulomb branch and $u$-direction with that of the mesons. In describing
the positions of NS branes, we need to fix the reference NS brane defined by $u=0$ in the $u-v$ plane. We choose
this to be the zeroth NS brane, and then we consider the relation of other NS branes w.r.t. the zeroth one. It is
most natural to consider the location of the first  NS brane in the $\CN=1$  as the moduli of the right-end
position of the D4 branes which were stretched between the zeroth and the first NS brane in the $\CN=2$ picture,
and the location of the second NS brane in the $\CN=1$ as that of the long D4 branes which were stretched between
the zeroth through the second NS brane. Hence the first NS brane is given by \bea u = W_1'(v) \eea and the second
NS brane is given by \bea u = W_1'(v) +W_2' (v),\eea and the intersection of the first and the second  NS branes is
given by $W_2'(v) =0$ and so its $v$-coordinate is exactly the expectation value of $\Phi_2$ when it is restricted
to the invariant subspace of $U(M_{22})$. Therefore, the D4 branes are located exactly at the intersection points
of the curves describing the NS branes (See Figure \ref{A3NS}).
Extending these arguments to the $\A_n$ quiver theory with the quadratic superpotential (\ref{supot}), we obtain
$(n+1)$ lines representing the NS branes, and thus there are $\frac{(n+1)n}{2}$ intersection points corresponding
to the stacks of  D4 branes. This is exactly the same as the number of the Higgsed gauge groups (\ref{Higgs}). For
the $\D_n$ case, the analysis is the same except the fact the last two lines are parallel (Figure \ref{dnNS}).

\begin{figure}
\centerline{\epsfxsize=120mm\epsfysize=70mm\epsffile{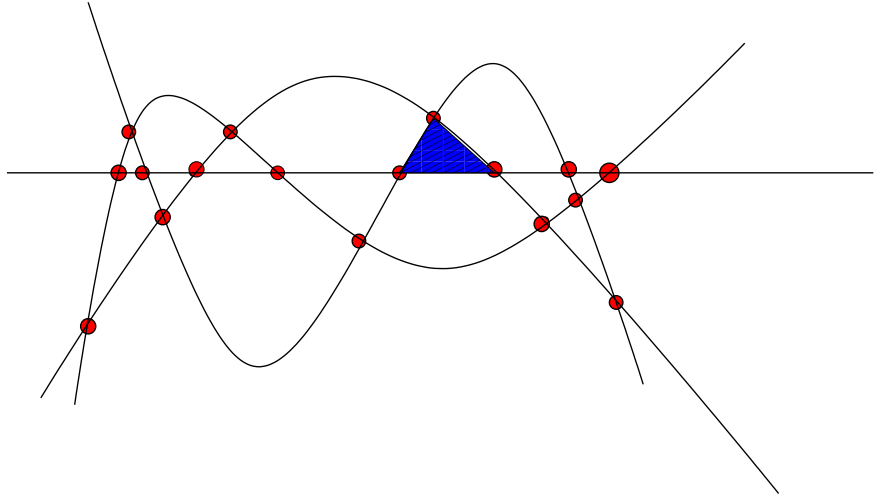}} \caption{A brane configuration for the $\CN=1$
$\A_3$ theory with $d_1 =2,~~d_{1,2} = 3$ and $d_{1,3} = 3$.}\label{NS}
\end{figure}

The general superpotential (\ref{supotg}) case is a straightforward extension.  The $l$-th NS brane is described
as the moduli of the right-end position of the D4 brane connecting the zeroth and the $l$-th NS brane in the $\CN
=2$ theory and  it is given by \bea\label{NScurves} u = \sum_{i=1}^{l} W_i'(v), \eea and the $v$-coordinates of
its intersection with the first NS brane $u=0$ are the common expectation values $\Phi_1, \ldots ,\Phi_l$ when
they are restricted to the invariant subspace under $U(M_{1,l})$. More generally, the $v$-coordinates of the
intersection of the $(l-1)$-th and the $m$-th NS branes are given by \bea W_{l}'(v) + W_{l+2}'(v) + \ldots +
W_m'(v) =0, \eea and hence they are the common expectation values of $\Phi_{l}, \ldots , \Phi_m$ on $U(M_{l,m})$.
The total number of intersection points is $d_{l,m}$ which agrees with the number of the Higgsed gauge groups. One
can give interpretation of the $u$-coordinates of the intersection points in terms of the expectation values of
the mesons. For $j=1$, (\ref{wjk}) means that the $v$-coordinates of the intersection of the first and the $k$-th
NS brane is given by $v_{1,k,l}$ and (\ref{bifundexp}) implies that the $u$-coordinates of the $m$-th curve \bea u
=\sum_{i=1}^m W_i'(v), \eea evaluated at $v= v_{1,k,l}$, will give the expectation values of the corresponding
mesons $Q_{m,m+1}Q_{m+1,m}$. Figure \ref{NS} shows a brane configuration for the $\CN=1$ $\A_3$ theory with 17
gauge groups Higgsed by a superpotential with $d_1 =2,~~d_{1,2} = 3$ and $d_{1,3} = 3$.
\section{Geometric Engineering for $\CN =1$  A-D-E Quiver Theory}
\pagestyle{myheadings}
\markboth{Duality and Confiement in $N=1$...}{K. Oh and R. Tatar}

 In this section, we describe Calabi-Yau
threefolds which are T-dual of the brane configurations of the $\CN =1$ theory. We first present the Calabi-Yau
threefolds and then we show that the T-dual picture is exactly the same as the brane configuration for the vacua.

The $\CN =1$ geometry for the $\A_n$ quiver theory with general superpotential $W$ (\ref{supotg}) is the minimal
resolution of a Calabi-Yau threefold defined in $\C^4$ by \bea\label{singandeg} X:\, xy - \prod_{p=0}^n
\left(u-\sum_{i=0}^p W_i'(v) \right) =0 \eea where $W_0'$ is defined to be zero. The singularities are isolated
and located at $x=y=0$ and the intersection of any two curves in the $u-v$ plane defined by \bea
\label{singularlocus} u= \sum_{i=1}^{j-1} W_i'(v),~~u=\sum_{i=1}^k W_i'(v).\eea

The singularities can be resolved by successive blow-ups which replace each singular point by a $\P^1$ cycle.
Therefore we see that
 the number of $\P^1$ cycles match the number of the Higgsed gauge groups. The
resolved space is covered by $(n+1)$ three dimensional complex spaces $U_p,
p=0, \ldots n$ with coordinates \bea \label{coord}
u_p = \frac{\prod_{j=0}^{p}\left(u- \sum_{i=0}^j W_i'(v)\right)}{x}, ~~x_{p} = \frac{x}{\prod_{j=0}^{p-1}\left(u-
\sum_{i=0}^j W_i'(v)\right)},~~ v_p =v,\eea where $x_0 =x$. They blow down to the singular threefold
(\ref{singandeg}) by
 \bea \label{sigma} \sigma :&&\tilde{X}:= U_0\sqcup U_2 \sqcup \ldots \sqcup U_n \to X,\\
&&U_p \ni (u_p, x_p, v_p) \mapsto \left\{
\begin{array}{l}x=\left\{ \begin{array}{l} x_0 \,\,\mbox{if}\,\, p=0,\\
x_p\prod_{j=0}^{p-1}\left( x_pu_p   +  \sum_{i=j+1}^p W_i'(v_p) \right) \,\,\mbox{otherwise}
\end{array}\right.
\\y= u_p  \prod_{j=p+1}^{n}\left( x_pu_p - \sum_{i=p+1}^j W_i'(v_p) \right)
 \\u=x_pu_p +  \sum_{i=0}^p
W_i'(v_p)  \\v=v_p
\end{array} \right. \eea
where $U_p \sqcup U_{p+1}$ means that the three spaces $U_p, U_{p+1}$ are glued together by \bea \label{gluedata}
x_{p+1} = u_p^{-1}, \,\, v_{p+1}=v_p,\,\, u_{p+1} = x_p u_p^2 - W'_{p+1}(v_p)u_p. \eea Thus the complex lines
$\C^1$ defined by  \bea \label{p1} W'_{p+1}(v_p)=0, ~~x_p=0 \eea in $U_p$  together with the complex lines $\C^1$
in $U_{p+1}$ defined by \bea \label{p2} W'_{p+1}(v_{p+1}) =0,~~u_{p+1} =0\eea  form the $\P^1$ cycles and there
are no other $\P^1$ cycles in  $U_p \sqcup U_{p+1}$. This is a generalization of the $\A_1$ quiver theory
considered in \cite{civ}. While the $\P^1$ in the resolution (\ref{anresol}) of the $\A_n$ singularity can move
freely in the $v$-direction, the above $\P^1$ cycles are frozen at \bea W_{p+1}'(v) = 0. \eea Hence the
supersymmetry is broken from $\CN=2$ to $\CN=1$.

\begin{figure}
\centerline{\epsfxsize=100mm\epsfysize=70mm\epsffile{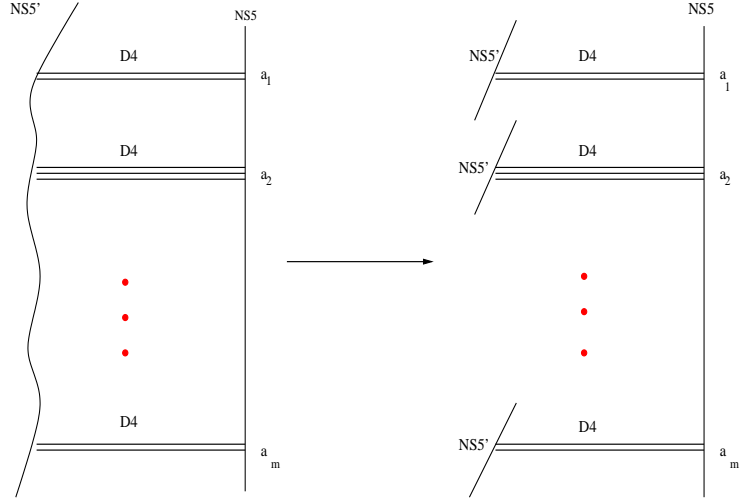}} \caption{The
$g_{m+1} \to \infty$ limit of $\A_1$ brane configuration with a
non-degenerate superpotential.} \label{CIV}
\end{figure}

\begin{figure}
\centerline{\epsfxsize=100mm\epsfysize=70mm\epsffile{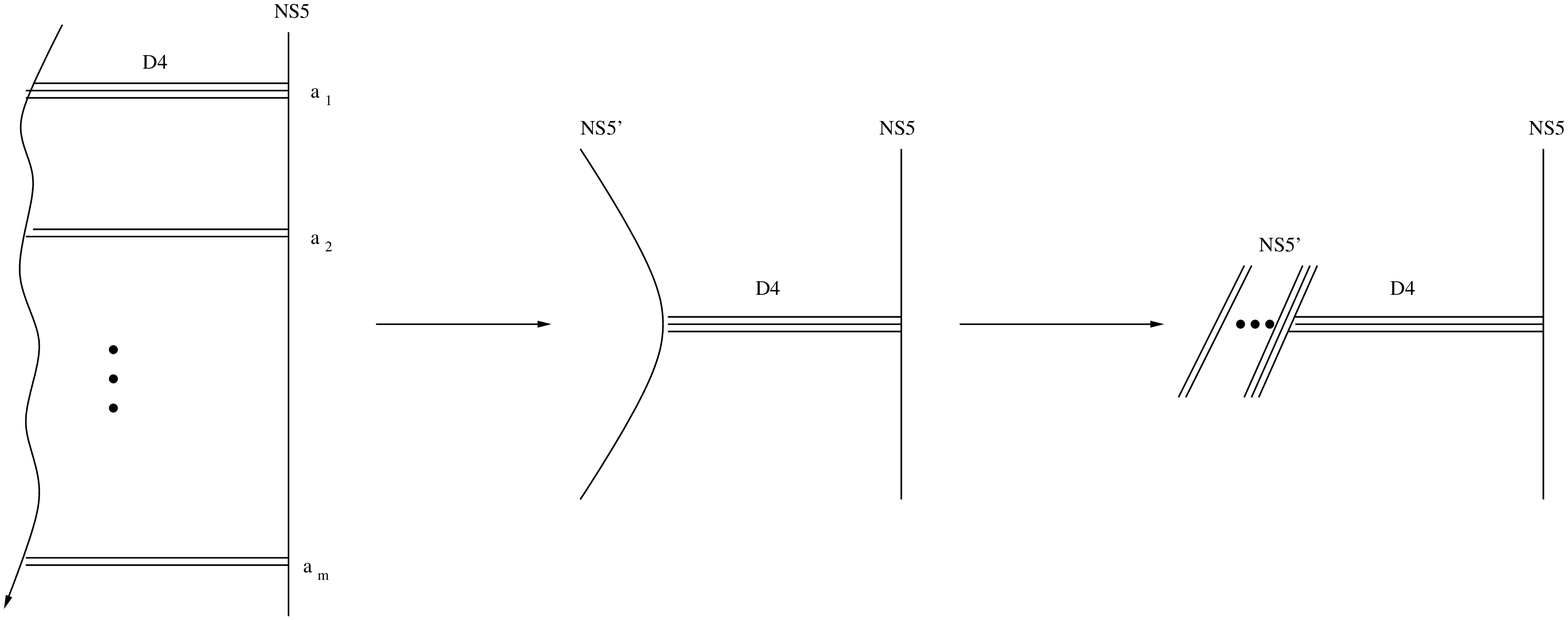}}
\caption{The transition $\A_1$ brane configuration with a
non-degenerate superpotential to a degenerate superpotential and the
$g_{m+1} \to \infty$ limit.} \label{manyNS}
\end{figure}

To study this phenomenon in details, consider the extreme case where \bea \label{degeneratesup} W_{p+1 }(v) =
\frac{g_{m+1}}{m+1}v^{m+1},~~m>1.\eea  Then the condition (\ref{gluedata}) becomes \bea \label{gluedata2} x_{p+1}
= u_{p}^{-1}, \,\, v_{p+1}=v_p,\,\, u_{p+1} = x_p u_p^2 - v_p^m u_p. \eea Locally in the neighborhood $U_{p}
\sqcup U_{p+1}$, this can be considered as the limit case of the $\A_1$ quiver theory considered in \cite{civ}. In
\cite{civ}, the $\CN=2$ theory has been deformed by a tree-level superpotential \bea \label{supotnd}
W_{\mbox{CIV}} = \sum_{p=1}^{m+1} \frac{g_p}{p} \Tr \Phi^p.\eea Then the classical vacua are located at \bea
W_{\mbox{CIV}}'(v) = \sum_{p=0}^{m} g_{p+1} v^p = g_{m+1} \prod_{p=1}^{d_k} (v -a_p). \eea While the D5 branes
wrapping $\P^1$ in the $\CN=2$ geometry can move freely along the $v$-direction, the D5 branes wrapped on $\P^1$
cycles in the $\CN =1$ geometry will be fixed at the vacua i.e. $v=a_p$ and the $\CN =1$ geometry can be described
as a union of two $\C^3$'s with the patching condition: \bea \label{civgluedata}x'= u^{-1}, ~~ v'=v,~~ u' = xu^2
+g_{m+1} \prod_{p=1}^{m} (v -a_p)u, \eea where $(x',v',u'),~~ (u,v, x)$ are coordinates systems for two $\C^3$'s
respectively. As we will see,  the T-dual picture is a brane configuration with $p$ stacks of D4 branes between a
straight NS and a curved NS brane of degree $m$  as in the far-left of
Figure \ref{manyNS} or \ref{CIV}.
\\

We may consider two limit of this configuration. If we take the limit where $g_{m+1} \to \infty$ while keeping the
ratio $g_i/g_{m+1}$ finite for $i=1, \ldots, m$, then the curved NS brane will break into $m$ lines and all of
them are separated from each other and  completely rotated so that the corresponding adjoint masses are infinity
in this limit. The configuration is shown in the right of Figure
\ref{CIV}.
\\
The $\CN =1$ geometry will be
given by the resolution of \bea xy + u \prod_{p=1}^m (v- a_p) =0, \eea  because the curve $u-g_{m+1}\prod_{p=1}^m
(v- a_p) =0$ will approach $\prod_{p=1}^m (v- a_p) =0$ in the limit.

On the other hand, if we take the limit where $g_i \to 0 $ for $i=1,
\ldots, m$, then all $a_i$ become zero and the NS brane is curved so that
it intersect with the straight NS branes in the $u-v$ plane with high
multiplicities which is shown in  the center of Figure \ref{manyNS}.
In the limit $g_{m+1} \to \infty$, the curved NS brane  breaks into $m$ lines and these lines will be on top of each other as in the far-right
of Figure \ref{manyNS}. Then  the gluing data (\ref{civgluedata}) becomes \bea \label{gluedata3} x'= u^{-1}, ~~
v'=v,~~ u' = xu^2,  \pmod {\CI^2}, \eea where $\CI$ is the ideal sheaf defining $\P^1$, the normal bundle $\CN =
\CI /\CI^2$ of $\P^1$ is $\CO(-2) + \CO$ so that the first order deformation space $H^0(\P^1, \CN) $ is one
dimensional, but there is an obstruction to deform this curve in the $m$-th order because the relation
(\ref{gluedata3}) breaks down in the $m$-th neighborhood $\CI^{m+1}$~\cite{reid}. As we observe previously, this
is due to the superpotential term $W_{CIV}(\Phi) = g_{m+1}\Phi^{m+1}$ \cite{bdlr, kklm}. This superpotential  can
be also generated by the holomorphic Chern-Simons actions~\cite{kklm,wittfloer,vafambundles}.

More generally, recall that the $\CN=1$ geometry for the degenerate superpotential is given by: \bea
\label{gluedata2p} x_{p+1} = u_{p}^{-1}, \,\, v_{p+1}=v_p,\,\, u_{p+1} = x_p u_p^2 - v_p^m u_p. \eea Let $E =
\cup_{i=1}^n E_i $ be the exceptional locus of the resolution $\sigma: \tilde{X} \to X$  where $E_i$ is isomorphic
to $\P^1$. Let $\CI_E$ be the ideal sheaf of $E$. The relation (\ref{gluedata2p}) is reduced to \bea
\label{normalbundle} x_{p+1} = u_{p}^{-1}, \,\, v_{p+1}=v_p,\,\, u_{p+1} = x_p u_p^2,~~ \pmod{\CI_E^2}, \eea which
can be identified with a tubular neighborhood of the total space of the normal bundle  $\CN_E = \CI_E/ \CI_E^2$
and $E$ can be identified with the zero section. The space (\ref{gluedata2p}) (more generally (\ref{gluedata}))
can be obtained as a modification of the complex structure from (\ref{normalbundle}). The modification of complex
structure is realized by perturbing the Cauchy-Riemann operator $\bar{\partial_{\bar{j}}}$ on (\ref{gluedata2p})
by \bea \bar{D} = \bar{\partial}_{\bar{j}} + A^i_{\bar{j}}
\partial_i \eea where $A^i_{\bar{j}}$ is an anti-holomorphic one form taking values in the tangent bundle of
(\ref{gluedata2p}). We assume that the zero section $E$ of the normal bundle $\CN_E$ remains holomorphic. The
space of $C^\infty$ deformations of $E$ is identified  with the space  $(s_1, s_2)$ of  $C^\infty$ sections of the
normal bundle $\CN_E$. The relevant holomorphic Chern-Simon action is \bea\label{chern-simon} \int_E (s_1 \bar{D}
s_2 - s_2 \bar{D} s_1). \eea More generally, we may consider the normal bundle  $\CN_{E_{j,k}}$ of \bea
E_{j,k}:=E_j\cup E_{j+1} \cup \cdots \cup E_k\eea and restrict the Chern-Simon action. Then the variations with
respect to the section  $(s_1, s_2)$ of $\CN_{E_{j,k}}$ give the conditions that the corresponding curve is
holomorphic.

So far we have counted the $\P^1$ cycles corresponding to the Higgsed branch whose gauge group is of the form
$U(M_{i,i})$ i.e. ones with the zero meson expectation values. There should be more $\P^1$ cycles  corresponding
to the non-zero meson expectation values. Where are they? They are not confined in a union of two open sets as
$\P^1$ cycles consider above, and rather spread out into several open sets. Consider the inverse images of the
singular points $p_{j,k,l}$ of $X$, which is one of the intersection points given by (\ref{singularlocus}), under
the blowing-down map $\sigma$ (\ref{sigma}). Then the inverse image $\sigma^{-1} (p_{j,k,l})$ in each open set
$U_p, ~~p=j, \ldots ,k$ will be a non-compact curve isomorphic to $\C^1$ and that they will glue together to form
a $\P^1$.

To see the exceptional $\P^1$'s corresponding to the non-zero meson expectation values more manifestly, we
introduce another resolution picture of (\ref{singandeg}) which is essentially the same as the one considered
above.  Consider a subvariety \bea\label{resAn} \tilde{Y}\subset \underbrace{\P^1 \times \cdots \times \P^1
}_{n}\times X \eea defined by for $p=1, \ldots, n$ \bea  &s_p x = t_p\prod_{j=0}^{p-1}\left(u- \sum_{i=0}^j
W_i'(v)\right),~~ s_p \prod_{j=p}^n \left(u- \sum_{i=0}^j
W_i'(v)\right) = t_p y,\nonumber \\
&t_{p-1}s_p =
 s_{p-1}t_p \left(u- \sum_{i=0}^{p-1}
W_i'(v)\right), \eea where $(s_p, t_p)$ is the homogeneous coordinates of the $p$-th $\P^1$ and $(s_0,t_0)$ is set
to be $(0,0)$.  Then the natural projection \bea \label{Anresol} \tau: \tilde{Y} \to X \eea will be the resolution
of $X$. It is easy to check that $\tilde{Y}$ is smooth. Consider a singular point given by $x=y=0$ and the
intersection of two curves (\ref{singularlocus}) with $j<k$.  The fiber of $\tau$ (\ref{Anresol}) over such a
point is a $\P^1$ given
by \bea \label{exceptional} \begin{array}{ll}t_p = 0 & \mbox{for $p< j$},\\
t_{p-1}s_p =s_{p-1}t_p \sum_{i=l}^{p-1}W_{i}'(v) &\mbox{for $j<p \leq k$},\\
 s_p = 0&\mbox{for $p> k$}.\end{array}\eea  Notice that $\sum_{i=l}^{p-1}W_{i}'(v)$ is non-zero
 for $j<p\leq k$. This shows that $\P^1$ is diagonally embedded into the product of $k-j+1$ $\P^1$'s beginning at
the $j$-th and ending in $k$-th $\P^1$ in (\ref{resAn}) which corresponds to the fact that the gauge groups
$U(M_{j,k,l}),~~l=1, \ldots , d_{j,k}$ are diagonally embedded in $U(N_j) \times \cdots \times U(N_k)$.

To see the T-dual picture, note that the circle action (\ref{anaction}) can be extended to $X$ and
lifted to $\tilde{X}$ as follows: \bea S_{X}:& &X \times \S^1 \to X,\\
&&\nonumber  (e^{i\theta}, x) = e^{i\theta}x, ~~(e^{i\theta}, y) = e^{-i\theta}y,~~(e^{i\theta}, u) =
u,~~(e^{i\theta}, v) = v,\\
S_{\tilde{X}}:&&\tilde{ X} \times \S^1 \to \tilde{X},\\
&& \nonumber (e^{i\theta}, u_p) = e^{i\theta}u_p, ~~(e^{i\theta}, x_p) = e^{-i\theta}x_p,~~(e^{i\theta}, v_p) =
v_p.\eea Under this action, the blowing-up map $\sigma$ (\ref{sigma}) and the gluing map (\ref{gluedata}) are
equivariant. In particular, the $\P^1$ cycles defined by (\ref{p1}) and (\ref{p2}) are invariant under the action.
So if we take T-dual along the orbits of the circle action $S_{\tilde{X}}$, then D5 branes wrapping $\P^1$'s will
become D4 branes on the interval and NS branes will appear where the orbits degenerate~\cite{bvs}. So there will
be one NS brane for each open set $U_p$ located at $u_p = x_p =0$ and stretched along $v_p$. Under the blow-down
map, it will map to \bea x=y=0,~~u = \sum_{i=0}^p W_i'(v).  \eea which shows NS brane is curved  into the
$u$-direction. So we will have  NS branes wrapping $(n+1)$ holomorphic curves in the $u-v$ space and the stacks of
D4 branes between them. This proves that the T-dual picture of D5 branes wrapping $\P^1$ cycles in $\tilde{X}$ is
exactly the same as the brane configuration constructed for the $\CN =1$ $\A_n$ quiver theory for the arbitrary
superpotential $W$ (\ref{supotg}).

The geometry for the $\CN=1$ $\D_n$ quiver theory is similar. It will be given by the resolution of a singular
threefold defined by the same equation as in $(\ref{singandeg})$ except the fact that the mass of the last two
adjoints are assumed to be the same i.e. $g_{n-1} =g_n$. So the analysis is similar.

\section{Large N Duality Proposal and Normalizable Deformations}
In the IR limit, the $\CN =1$ A-D-E quiver theory is equivalent to a pure $\CN=1$ gauge theory. So we expect to
have gaugino condensation  and mass gap as noticed in  \cite{ckv}. In the large N description, the theory lives on
a geometry where  the $\P^1$ cycles have shrunk and $\S^3$ cycles have grown and RR fluxes through them and NS
fluxes through their dual cycles have been created. We will mainly consider the case when $W_i$'s are of the same
degree.

We first consider the case of the degenerate superpotential (\ref{supotd}) whose $\CN=1$ geometry is
:\bea\label{singandegd} X_{\mbox{deg}}:\, F(x,y,u, v) :=xy - \prod_{k=0}^n \left(u- \sum_{i=0}^k g_i v^m\right)
=0, \eea where $g_0$ is defined to be zero.

The singularity $X_{\mbox{deg}}$ admits the $U(1)$ symmetry group \bea x \to e^{i\theta/2} x,~~y\to
e^{i\theta/2}y,~~u\to e^{i\theta/(n+1)}u,~~v \to e^{i\theta/m(n+1)} v.\eea If $F$ is viewed as the superpotential
of  a Landau-Ginzburg theory~\cite{Gukov}, it would flow to a superconformal theory with central charge given by
\bea \hat{c} = (1-2Q(x)) + (1 -2Q(y)) + (1-2Q(u)) +(1-2Q(v)) = \frac{2mn-2}{m(n+1)}. \eea The miniversal
deformation space~\cite{AGV} of the singularity, which describes the most general complex deformations, is given
by the chiral (or Milnor)  ring  \bea \CR :=\frac {\C \{ x, y, u,v \}}{\left(\partial F/\partial x,
\partial F/\partial y,\partial F/\partial u,\partial F/\partial v\right)}, \eea
which  is generated by the monomials $m_{i,j}= u^iv^j$ with the charges \bea Q_{i,j}= \frac{i}{n+1} +
\frac{j}{m(n+1)}. \eea The Poincar\'e series  is given by \bea P_{\CR}(t) := \sum_{m_{i,j} \in \CR} t^{Q_{i,j}} =
\frac{(1-t^{1-Q(x)})(1- t^{1-Q(y)})(1- t^{1-Q(u)})(1-
t^{1-Q(v)})}{(1-t^{Q(x)})(1-t^{Q(y)})(1-t^{Q(u)})(1-t^{Q(v)})},\eea and so the dimension of the ring $\CR$ is \bea
\mbox {dim}_\C \CR = P_{\CR}(1)= n(nm+m-1).\eea The Milnor ring characterizes the geometry of the generic
deformation of $F =0.$ Namely the Milnor fiber \bea \label{Milnorfiber}\{(x,y,u,v) \in \C^4 | F (x,y,u,v) =
\epsilon \ll 1\} \eea is homotopic to a bouquet of $n(nm+m-1)$ $\S^3$'s. Moreover the most general deformation of
$F(x,y,u,v)$ is given by \bea \label{defdegan} X_\ep:F_{\ep}(x,y,u,v):=F(x,y,u,v) + \sum_{m_{i,j}\in \CR}
\ep_{i,j} m_{i,j}=0.
\eea As shown in \cite{Gukov}, in order for the deformation to corresponds to the dynamical parts of the theory at
the singularity, the cohomology classes created by the deformation should be supported on the singularity i.e.
vanishing cohomologies.  The Poincar\'e duals to the vanishing cohomology classes are the vanishing homologies
which arises from quadratic singularities (i.e. conifold singularities). These classes correspond to the
normalizable (including log-normalizable) three forms which satisfies: \bea \label{normalizable} \lim_{\delta \to
0} \int_{X_\ep \cap \B^6_\delta} \left| \frac{\partial \Omega_{X_\ep}}{\partial \epsilon_{i,j}}\right|^2 \to
\infty, \eea where $\Omega_{X_\epsilon}$ is the holomorphic three form on $X_\ep$ and $\B^6_\delta$ is the six
dimensional ball with radius $\delta$ located at the singularity. This normalizability is necessary in order for
the deformation to describe the large N dual of the original theory because the original theory should be valid
near the singularity, and also insures that the geometric transitions will be the conifold transitions locally.

The $U(1)$ symmetry of the singularity $F=0$ can be extended to the deformed space $F_\ep =0$ by giving the
charges to the deformation parameters $\ep_{i,j}$ and further extended to $\wedge^3 T^*(F_\ep =0)$. The
holomorphic three form \bea \Omega_\ep = \frac{dy\wedge du \wedge dv }{\partial F_\ep /\partial x} \eea has charge
\bea Q(\Omega_\ep) = Q(x) +Q(y) +Q(u) +Q(v) -1 = 1- \frac{\hat{c}}{2}. \eea The integral (\ref{normalizable}) can
be written as \bea \frac{\partial^2}{\partial \ep_{i,j}\overline{\partial} {\ep_{i,j}}} \int_{X_\ep \cap
\B^6_\delta} \Omega_{X_\ep}\wedge \overline{\Omega}_{X_\ep}, \eea and  it diverges when \bea  Q(\Omega_{X_\ep}) +
Q(u^iv^j)-1 \leq 0. \eea This holds when
 \bea
Q(u^iv^j) =  \frac{i}{n+1} + \frac{j}{m(n+1)} \leq \frac{\hat{c}}{2}=\frac{mn-1}{m(n+1)}.
 \eea
Therefore, the normalizable deformations are generated by the monomials $u^iv^j$ with $mi +j \leq mn-1$ and there
are $m(n+1)n/2$ of them which give the dynamical parts of the dual theory (gluino condensation). Since there are
$n(nm+m-1)$ deformation parameters in total, there are \bea \frac{m(n+1)n}{2} - n \eea non-normalizable
deformations which must be used to specify the theory externally (i.e. fixing the parameters of the tree-level
superpotential). So in the Milnor fiber (\ref{Milnorfiber}), there are two kinds of $\S^3$, ones corresponding to
the normalizable deformations and the others corresponding to the non-normalizable deformations.

We now further deform the theory by adding lower order terms in the superpotential i.e. we consider the general
superpotential (\ref{supotg}) with $d_i =m$. Then the $\CN =1$ geometry is given by the deformation  of : \bea
\label{newAn} X^m:\, F_m(x,y,u,v):= xy - \prod_{i=0}^n \left( u - \sum_{i=0}^k W_i'(v)  \right) =0,\eea which is
the same as (\ref{singandeg}) with constraint $d_i = \mbox{deg $(W_i)$} = m$ (by abusing the notation, we write
$X$ for $X^m$ if there is no confusion). Now $X$ has only conifold singularities (this is due to our assumption
that all the curves $u=\sum_{i=0}^k W_i'(v) $ are smooth and they meet transversally each other.)  and all
non-normalizable deformations are absorbed in the lower order terms of the superpotential and the remaining
deformations are all normalizable. Hence the deformation of $X$ is given by \bea X_{\mbox{def}}:\, G_m(x,y,u,v)
:= F_m(x,y,u,v) + \sum_{mi+j \leq mn-1} \ep_{i,j} u^i v^j=0. \eea Therefore the geometric transition is the
transition from the resolution $\tilde{X}$ of $X$ to the deformation $X_{\mbox{def}}$ where the exceptional
$\P^1$ cycles have been replaced by the Lagrangean three cycles $\S^3$'s.

Before we discuss the geometric transition in full details, we would like to investigate the deformation from the
$\CN=1$ geometry $X_{\mbox{deg}}$ with the degenerate superpotential $\Wd$ to the $\CN=1$ geometry $X$ with the
general superpotential $W$. In deforming from $X_{\mbox{deg}}$ to $X^m$, there will be  $ m(n+1)n/2 -n$ $\S^3$'s
appearing in the deformed geometry $X^m$ corresponding to the non-normalizable deformations. The way the $\S^3$
cycles  appear can be seen as follows. There are $m(n+1)n/2 -n$ non-contractible 1-cycles in the NS brane
configuration in the $u-v$ plane defined by (\ref{NScurves}). Figure \ref{NS} has 14 non-contractible 1-cycles. In
general, one can prove that there are \bea \label{onecycles} d -n = \sum_{j=1}^n \sum_{k=j}^n d_{j,k} - n
=\sum_{k=1}^n \sum_{j=1}^k d_{j,k} -n \eea non-contractible 1-cycles by induction on $n$. For $n=1$, the first NS
brane $u= W_1' (v)$ meets the $v$-axis at $d_{1,1}$ points so that there are $d_{1,1} -1$ enclosed 1-cycles since
any two consecutive intersection  points determine 1-cycle. The $k$-th NS brane meets the $(j-1)$-th NS brane at
$d_{j.k}$ points and so there are $\sum_{j=1}^{k} d_{j,k}$ new intersection points comparing with the case
$n=j-1$. This produces $\sum_{j=1}^{k} d_{j,k}-1$ new non-contractible 1-cycles. Hence this proves
(\ref{onecycles}).

\begin{figure}
 \centerline{\epsfxsize=120mm\epsfysize=70mm\epsffile{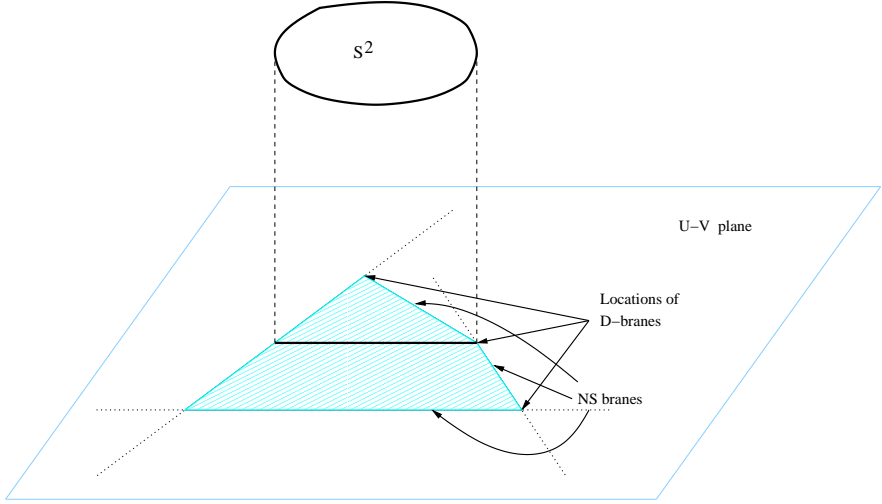}}
\label{Geom}\caption{The threefold geometry over $u-v$ plane:  An $\S^2$ lies over the thick line and an $\S^3$
lies over the shaded quadrilateral.}
\end{figure}

Now note that each 1-cycle $\gamma$ bounds a compact real two dimensional domain $\Delta_\gamma$ in the $u-v$
plane.  For example, one of them in the $\A_3$ quiver theory is shown in Figure \ref{NS} as a shaded region.
 We claim that the inverse image of the compact domain bounded by the
1-cycle under the projection map \bea \label{projection} \pi: X^m \to \C^2:\,\,(x,y,u,v) \to (u,v). \eea contains
a $\S^3$. One can see this heuristically as follows: the inverse image of the interior point of the domain is a
hyperboloid i.e. $\C^*$ and at the boundary point of the domain, the waists will shrink to zero i.e. the inverse
image of a union of two complex lines meeting transversely. In Figure \ref{Geom}, the inverse image, up to
homotopy, of the line joining the boundary points of the polyhedron formed by the NS branes in the case of the
quadratic superpotential is shown, which is homotopic to $\S^2$.
More rigorously, we may assume, after coordinate changes, that the domain $\Delta_\gamma$ lies in the real part of
the $u-v$ plane and $X^m$ is defined by an equation \bea x^2 +y^2 + \prod_{i=0}^n \left( u - \sum_{i=0}^k W_i'(v)
\right)=0. \eea  and the function \bea f(u,v) = \prod_{i=0}^n \left( u - \sum_{i=0}^k W_i'(v) \right) \eea as a
function of the real variables is not positive. Note that $f(u,v)$ is zero on the boundary of each domain.
Moreover, one can show that the function $f(u,v)$ has a unique extremum in the interior of each domain bounded by
1-cycles by induction on $n$.  Then one can construct a homeomorphism $h$ from a unit disk $D$ in the real $u-v$
plane to a domain bounded by 1-cycle sending each concentric circles to the isothermal curves of $f(u,v)$. By
pulling back the fiberation \bea \pi_{\Delta_\gamma} : \pi^{-1} (\Delta_\gamma) \to \Delta_\gamma \eea  to the
disk  $D$, we obtain a fiberation $h^*(\pi_{\Delta_\gamma})$ over a unit disk $D$ which can be written as \bea
h^*(\pi) : \{(x,y,u,v) \in \R^4 | x^2 +y^2 +u^2 +v^2 =1\} \to D,~~ (x,y,u,v) \to (u,v) \eea and this shows that
$\pi^{-1}(\Delta_\gamma)$ is homeomorphic to $S^3$.

So there are $mn(n+1)/2 -n$ (in general $d-n$) $\S^3$ cycles in the blown-down geometry  $X^m$ (in general $X$)
which are non-dynamical parameters of the theory. Of course, the appearing $\S^3$'s  are completely fixed by the
superpotential so by the expectation values of the adjoints and bifundamentals. Conversely, the superpotential is
fixed by the $\S^3$'s. Having all non-dynamical deformations fixed by expectation values of the adjoints and
bifundamentals (in geometry, there are only conifold singularities in the blown-down geometry), the $\CN =1$
geometry (\ref{sigma}) can go through the geometric transition where the rigid $\P^1$'s will disappear and will be
replaced by the finite size of $\S^3$. The equation of the new geometry $Y_\ep$ is given by \bea
\label{Yep}Y_\ep:~ G_m(x,y,u,v)=xy - \prod_{i=0}^n \left( u - \sum_{i=0}^k W_i'(v) \right)  + \hspace*{-.1in} \sum_{mi+j \leq
mn-1} \hspace*{-.1in} \ep_{i,j} u^i v^j=0. \eea Under this deformation, each singular point of the blown-down geometry $X$ which
corresponds to the intersection points of the NS branes (Figure \ref{A3NS})  split into two points on the smooth
curve in the $u-v$ plane  and replaced by $\S^3$ cycles. The holomorphic three form $\Omega$ on $Y_\ep$ is given
by \bea \Omega = \frac{dx dy du dv}{d G_\ep} = \frac{dy du dv}{y}.\eea  Since the NS branes meet transversally at
each singular point $p_{j,k,l}$, we can locally write the equation of $X$ as $ x^2 +y^2 +u^2 +v^2 =0 $ and, thus
the  deformed geometry $Y_\ep$ can be written as \bea x^2 + y^2 + u^2 + v^2 = \mu_{j,k,l}
 \eea with holomorphic three form
 \bea
 \Omega = \frac{dx\, dy\, du}{\sqrt {\mu_{j,k,l} - x^2 -y^2 -u^2}}
 \eea
Therefore, the period of the holomorphic three-form $\Omega_\ep$ over the 3-cycle $A_{j,k,l}$ of (\ref{Yep}),
which is a compact 3-sphere, is given by \bea S_{j,k,l} = \int_{A_{j,k,l}} \Omega_\ep \sim \frac{\mu_{j,k,l}}{4}
\eea and the period over the dual $B_{j,k,l}$ cycle is \bea \Pi_{j,k,l} =\int_{B_{j,k,l}}^{\La_{j,k,l}} \Omega_\ep
\sim \frac{1}{2\pi i} \left( -3S_{j,k,l} \log \La_{j,k,l} - S + S \log S\right) + \ldots. \eea As in the other
geometric transitions ~\cite{civ, ckv}, $S_{j,k,l}$ is identified with the glueball field $S_{j,k,l} =
-\frac{1}{32 \pi^2} \mbox{Tr}_{SU(M_{j,k,l})} W_\alpha W^\alpha $ of the non-Abelian factor $SU(M_{j,k,l})$  of
$U(M_{j,k,l})$ in (\ref{Higgs}) in the dual theory. The $S_{j,k}$ will be massive and obtain particular
expectation values due to the superpotential $\We$. The dual superpotential $\We$  arises from the non-zero fluxes
left after the transition. The deformed geometry will have $M_{j,k,l}$ units of $H_R$ flux through the $A_{j,k,l}
\cong \S^3$ cycle due to $M_{j,k,l}$ D5 branes wrapped on the $\P^1$ cycle before the transition, and there is
also an $H_{NS}$ flux $\alpha_{j,k}$ through each of the dual non-compact cycle $B_{j,k,l}$ with $2\pi i
\alpha_{j,k} = 8\pi^2/g_0^2$ given in terms of the bare coupling constant $g_0$ of the $U(M_{j,k,l})$ theory.

The appearance of an NS field in the deformed geometry can be understood as follows. In the mirror resolve
geometry it is an NS 4-form related to the change in complex structure \cite{vafa} (see also \cite{lust}). This
should be mapped in the the resolved geometry to  $H_{NS}$ which is due to a change in K\"ahler structure of
$Y_\ep$ whose origin comes from the K\"ahler structure change of $\tilde{X}$  due to the superpotential term \bea
\sum_{i=j}^k W_i \eea whose leading coefficient gives the size of $\P^1$ arising by blowing up the singular point
$p_{j,k,l}$. So $H_{NS}$ is of the form \bea H_{NS} = \frac{1}{2} (\partial + \bar{\partial}) J \eea where $J$ is
the $(1,1)$ K\"ahler form representing the K\"ahler structure change of $Y_\ep$.

Thus the effective superpotential is \bea\label{supoteff} - \frac{1}{2\pi i} \We = \sum_{j=1}^n\sum_{k=j}^n
\sum_{l=1}^{d_{j,k}}(M_{j,k,l} \Pi_{j,k,l} + \alpha_{j,k} S_{j,k,l}). \eea After identifying $\La_{j,k,l}$ with
the UV-cutoff, we obtain the usual lower energy superpotential associated with the $SU(M_{j,k,l})$ glueballs:
\bea \We = \sum_{j=1}^n\sum_{k=j}^n S_{j,k,l}\left(\log \frac{\La_{j,k,l}^{3 M_{j,k,l}}}{S_{j,k,l}^{M_{j,k,l}}} +
{M_{j,k,l}}\right). \eea Integrating out the massive $S_{j,k,l}$ by solving \bea \frac{\partial \We}{\partial
S_{j,k,l}} =0 \eea leads to $M_{j,k,l}$ supersymmetric vacua of $SU(M_{j,k,l})$ $\CN =1$ supersymmetric
Yang-Mills: \bea \label{Sjkl} \ev{S_{j,k,l}} = \exp (2\pi i m/M_{j,k,l}) \La_{j,k,l}^3, ~~m=1, \ldots, M_{j,k,l}.
\eea

The dual theory obtained after the transition is an $\CN =2$ \bea\prod_{j=1}^n\prod_{k=j}^n\prod_{l=1}^{d_{j,k}}
U(1) \equiv U(1)^{d}\eea gauge theory broken to $\CN=1$ $U(1)^{d}$ by the superpotential $\We$ (\ref{supoteff})~\cite{tv}.
The $S_{j,k,l}$, which are the same $\CN=2$ multiplet as the the $U(1)^d$, get masses and frozen to particular
$\ev {S_i}$ by $\We$. On the other hand, the $\CN=1$ $U(1)^d$ gauge fields remain massless. The couplings
$\tau_{j,k,l, j',k',l'}$ of these $U(1)$'s can be determined by $\Pi_{j,k,l}$ or the $\CN =2$ prepotential
$\CF(S_{j,k,l})$, with $\Pi_{j,k,l} = \partial \CF /\partial S_{j,k,l}$ : \bea \tau_{i,j,k, i',j',k'} =
\frac{\partial \Pi_{i,j,k}}{\partial S_{j',k',l'}} = \frac{\partial^2 \CF (S_{j,k,l})}{\partial S_{j,k,l}\partial
S_{j',k',l'}}.\eea The couplings should be evaluated at the vacua $\ev {S_{j,k,l}}$ obtained in (\ref{Sjkl}). As
in \cite{civ}, the coupling constants of the $U(1)$ factors are related to the period matrix of the curve \bea
\label{couplingcurve} \prod_{i=0}^n \left( u - \sum_{i=0}^k W_i'(v) \right)  + \sum_{mi+j \leq mn-1} \ep_{i,j} u^i
v^j=0, \eea which is an $(n+1)$-fold covering of the $v$-space. By integrating the period integral of $Y_\ep$
first along the fibers of the projection from $Y_\ep$ to the $v$-space, one might be able to compute the coupling
constants $\tau_{j,k,l,j'k'l'}$ in terms of the periods integral  of the curve (\ref{couplingcurve}). Instead of
doing this, we will take type IIA picture by T-duality and then lift to M-theory. From M-theory perspective, it
will be clear that  the curve (\ref{couplingcurve}) is indeed a Seiberg-Witten curve for the $\CN=1$ $U(1)^d$
theory.

To consider type IIA picture, note that the circle action considered for the various geometries so far can also be
extended to the geometry after the transition because it acts only on the $x-y$ space. The orbits degenerate along
$x=y=0$. So in the T-dual picture, we have an NS brane wrapping the curve in the $u-v$ plane defined by \bea
\prod_{i=0}^n \left( u - \sum_{i=0}^k W_i'(v) \right)  + \sum_{mi+j \leq mn-1} \ep_{i,j} u^i v^j=0, \eea which is
exactly (\ref{couplingcurve}).

\section{M Theory and the M5 brane Transition}

We now lift the type IIA brane configurations for the $\CN=1$ $\A_n$ quiver theory, deformed by the superpotential
(\ref{supot}), to M theory and investigate the large $N$ limit via Witten's MQCD formalism. For simplicity, we
will often restrict to the case with the degree of $W_i$ are the same, denoted by  $m+1$. We denote the finite
direction of D4 branes by $x^7$ and the angular coordinate of the circle $\S^1$ in the 11-th dimension
 by $x^{10}$. Thus the NS branes are separated along the $x^7$
direction. We combine them into a complex coordinate \bea t = \exp ( -R^{-1}x^7 - i x^{10})\eea where $R$ is the
radius of the circle $\S^1$ in the 11-th dimension. In MQCD \cite{witqcd}, the classical type IIA brane
configuration turns into a single fivebrane whose world-volume is a product of the Minkowski space $\R^{1,3}$ and
the M-theory curve $\Sigma$ in a flat Calabi-Yau manifold \bea \label{M}M = \C^2 \times \C^*,\eea whose
coordinates are $u,v,t$.

\begin{figure}
\centerline{\epsfxsize=100mm\epsfysize=50mm\epsffile{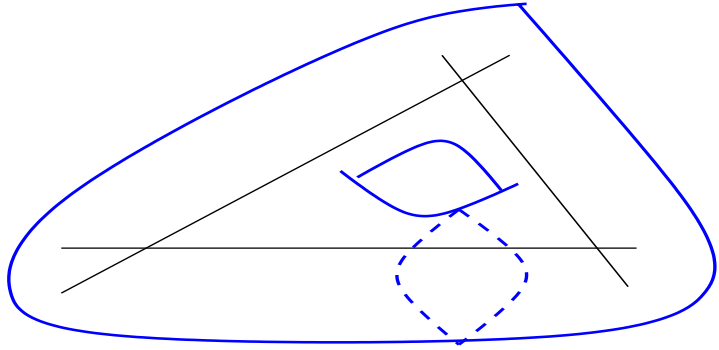}}  \caption{The line configuration has deformed to
a smooth  curve by quantum effect in M-theory.} \label{SWNS}
\end{figure}

Classically , the $u, v$-coordinates of the M-theory curve $\Sigma$ describes the location of NS branes in the
type IIA picture. Hence it is given by Figure \ref{NS}: \bea \label{nsclassic}\prod_{i=0}^n \left( u -
\sum_{i=0}^k W_i'(v) \right) =0. \eea By lifting this to the M-theory, the location of the D4 branes on the NS
brane will be smeared out as D4 brane acquire 11-th direction. So, at quantum level, the configuration
(\ref{nsclassic}) will be deformed and the $u, v$ coordinates of $\Sigma$ will satisfy: \bea \label{SigmaNS}
 \prod_{i=0}^n \left( u - \sum_{i=0}^k W_i'(v)  \right) + \sum_{mi+j \leq mn-1} \ep_{i,j} u^i v^j=0.
\eea Note that we have added the only deformations which will modify the intersection of NS5 and D4 branes in type
IIA picture and  there are $m(n+1)n/2$ complex parameters corresponding to the number of the stacks of D4 branes.
Figure \ref{SWNS} shows the quantum deformation of the NS brane configuration for the $\A_2$ theory with quadratic
superpotential (Figure \ref{A3NS}) where the union of three lines is deformed into a curve of genus one.
\\
\\

When $\Sigma$ is projected onto the $u-v$ plane, it will be denoted by $\Sigma_{NS}$ and becomes a
non-hyperelliptic curve for $n>1$, unlike most cases in the literature. In fact, $\Sigma_{NS}$  is a $(n+1)$-fold
branched covering of a complex plane and  can be compactified to a  projective curve $\overline{\Sigma_{NS}}$ with singularities at
infinity for $m>1$ with  geometric genus (i.e. the genus after resolving singularities) is \bea \label{gg}
g(\overline{\Sigma_{NS}}) = \frac{n(mn +m -2)}{2} \eea which is the same as the rank of the first homology of the NS brane
configurations in the $u-v$ plane. Heuristically, one can see from Figure \ref{SWNS} that $\Sigma_{NS}$ is a small
deformation of the NS brane configuration Figure \ref{A3NS}, so the number of holes (i.e. genus) is the same as
the number of non-contractible 1-cycles in the NS brane configuration. In general, the genus of $\overline{\Sigma_{NS}}$ is
\bea \label{Gg} g(\overline{\Sigma_{NS}}) = d-n \eea where $d$ is the number of the Higgsed groups defined in (\ref{d}). For
example, $\Sigma_{NS}$ corresponding to Figure \ref{NS}  will be a curve of genus 14.
As explained in \cite{dot1}, the coupling of over-all $U(1)$ of each $U(N_i)$ is described by the non-compact part of the Jacobian $J(\Sigma_{NS})$ which fits into an exact sequence of algebraic groups:
\bea
1 \longrightarrow (\C^*)^n \longrightarrow J(\Sigma_{NS}) \longrightarrow J(\overline{\Sigma_{NS}}) \longrightarrow 0.
\eea
Hence this produces the exact number of $U(1)$ gauge factors in the $\CN =1$ transition!

Because $\Sigma$ is not rational, one cannot parameterize the $t$-coordinates of $\Sigma$ in terms of $u$ or $v$.
Since there are $M_{j,k,l}$ D4 branes at each intersection point $p_{j,k,l},~~l=1, \ldots ,m$ of the NS branes
\bea u= \sum_{i=0}^{j-1} W'_i(v),~~u=\sum_{i=0}^{k} W'_i(v) \eea in the $u-v$ plane corresponding to the gauge
group $U(M_{jk})$ classically, $\Sigma$ is a deformation of a finite covering of the NS brane configuration
ramified over the intersection points $p_{j,k,l}$ and it is possible to give a parametric description locally
which will be used later.

As we go through the transition, the sizes of $\P^1$'s get smaller and there are gaugino condensations. In
M-theory the $t$-coordinates which describe the $SU(M_{j,k,l}$ parts of $U(M_{j,k,l})$ will be fixed and then the
remaining $U(1)$ parts of the theory  will be given by  a plane curve in the $u-v$ plane given by \bea
\label{GeomCurve} \prod_{i=0}^n \left( u - \sum_{i=0}^k W_i'(v) \right) + \sum_{mi+j \leq mn-1} \z_{i,j} u^i
v^j=0. \eea where $\z_{i,j}$ are fixed by the gaugino condensation which breaks the chiral symmetries. So this is
the quantum moduli of the remaining $U(1)^d$ theory where $d$ is the total number of the Higgsed gauge groups
given in (\ref{d}) since the $U(1)$ parts of $U(N)$ theory is parameterized by the center of mass coordinates of
D4 branes. Hence from the M-theory, the geometric transition is nothing but a transition from the space curve
$\Sigma$ to the plane curve.

Let us consider this dual theory from $\CN=2$ Seiberg-Witten theory. To motivate our method, we consider the
$\A_1$ case. The $\CN=2$ theory deformed by $W_{\mbox{tree}} = \sum_{i=1}^{m+1} g_r u_r $ has unbroken
supersymmetry only on submanifolds of the Coulomb branch, where there are additional massless fields besides the $u_r$.
The additional massless fields are the magnetic monopoles or dyons, which become massless on some particular
submanifolds $\ev{u_p}$~\cite{seibergwitten}. Near a point with $l$ massless monopoles, the superpotential is \bea
W= \sum_{k=1}^l M_k(u_r) q_k\tilde{q_k} + \sum_{p=1}^{m+1} g_p u_p, \eea and the supersymmetric vacua  are
$\ev{u_p}$ satisfying \bea M_k(\ev{u_p}) =0,~~\sum_{k=1}^l \frac{\partial M_k (\ev{u_p})}{\partial u_p}
\ev{q_k\tilde{q_k}} + g_p =0, \eea where the first equations are for all $k=1, ldots ,l$ and the second for all
$r=1,\ldots , N$ (with $g_p =0$ for $p >m+1$). The Seiberg-Witten curve of the $U(N)$ theory is \bea u^2 + P(v,
u_r) u + \Lambda^{2N} =0,~~ P(v, u_r) \equiv \det (v -\Phi) = \prod_{i=1}^N (v - e_i). \eea After the coordinate
change $2(u+P(v, u_r)/2)$ by $y$, the curve becomes \bea y^2 = P(x, u_r)^2 - 4 \Lambda^{2N}. \eea This is a
hyperelliptic curve which has $N$ branch cuts centered  about the eigenvalues $e_i$ with endpoints $e_i^{-}$ and
$e_i^{+}$.  The condition for having $N-n$ mutually local massless magnetic monopole  is that \bea P(x, \ev{u_p})
- 4 \Lambda^{2N} = (H_{N-n}(v))^2 F_{2n}(v), \eea where $H_{N-n}$ is polynomial in $v$ of  degree $N-n$ and
$F_{2n}$ is a polynomial in $x$ of degree $2n$.  Out of $N$ branch cuts, $N-n$ disappear and the curve
becomes singular at those points. The remaining $n$ massless photons is described by the reduced curve \bea y^2 =
F_{2n} (x, \ev{u_r}) = F_{2n}(x, g_p, \Lambda ). \eea Classically, we have \bea P(v, u_r) = \prod_{i=1}^n
(v-v_i)^{N_i} \eea where $v_i$ are the eigenvalues of $\Phi$ obtained by solving the tree-level superpotential.
Since $\ev{u_r}$ are obtained from the classical values after quantum corrections, we have \bea F_{2n} =
\prod_{i=1}^n (v -v_i^{+}) (v - v_i^{-}) \eea where $v_i^{\pm}$ are quantum corrections of $v_i$. After possible
change of the scale $\Lambda$, we can consider \bea y^2 = \prod_{i=1}^n (v-v_i)^2 - 4 \Lambda^{2n} \eea or \bea
u^2 +\prod_{i=1}^n (v-v_i) u +\Lambda^{2n} =0\eea as a reduced curve describing the remaining massless photons.

More generally we begin with the $\A_n$ brane configuration for the ${\cal N} = 2$ theory with the gauge group
$\prod_{i=1}^n U(N_i)$ and hypermultiplets in the representation $\sum_{i=1}^{n-1} ({\bf N}_i, \bar{{\bf
N}}_{i+1})$. The corresponding Seiberg-Witten curve will be a curve defined by a polynomial
$F(t,v)$~\cite{ENR,KMV}: \bea \label{N=2curve}u^{n+1} - P_1(v) u^n &+ \sum_{j=2}^{n} (-1)^j (\prod_{i=1}^{j -1}
\La_i^{(j-i)\be_i})P_j(v) u^{n+1-j} \nonumber \\
&+ (-1)^{n+1}(\prod_{i=1}^{n} \La_i^{(n+1-i)\be_i}) = 0, \eea where $\be_i =
2N_i- N_{i-1} -N_{i+1}$ is the $\be$-function coefficient for $U(N_i)$. Here $\Lambda_i$ is a $\CN =2$ QCD scale
for $U(N_i)$ theory and $F(u,v)$ is of degree $(n+1)$ in $u$ so that for each $v$ there are $(n+1)$ roots
corresponding to $(n+1)$ NS branes and $P_i$ is of degree $N_i$ whose zeroes are the positions of the $N_i$ D4
branes stretched between the $(i-1)$-th and $i$-th NS brane. Hence we have \bea P_i(v) = \det (v -\Phi_i) =
\prod_{k=1}^{N_i} (v - e_k^{(i)}), \eea where $e_k^{(i)}$ are eigenvalues of $\Phi_i$.  The curve
$(\ref{N=2curve})$ describes an $(n+1)$-fold branched covering of the complex plane with $(i-1)$ and $i$-th
sheets connected by $N_i$ branch cuts centered about the eigenvalues $e_k^{(i)}$  with endpoints $e_k^{(i)-}$ and
$e_k^{(i)+}$. According to Seiberg-Witten theory~\cite{seibergwitten}, the renormalized order parameters, their
duals  and the prepotential $\CF$ are given by \bea  a_k^{(i)} =\frac{1}{2 \pi} \oint_{A_k^{(i)}} \lambda,~~~
a_{D,k}^{(i)} =\frac{1}{2 \pi} \oint_{B_k^{(i)}} \lambda,~~a_{D,k}^{(i)} = \frac{\partial \CF}{\partial
a_k^{(i)}}, \eea where $\lambda$ is the Seiberg-Witten differential, and $A_k^{(i)}$ and $B_k^{(i)}$ are a set of
canonical homology cycles for the curve (\ref{N=2curve}). The cycle $A_k^{(i)}$ is chosen to be simple contour on
sheet $i$ enclosing the branch cut centered about $e_k^{(i)}$ and $B_k^{(i)}$ is the dual cycle. The gauge
symmetry is generically broken to \bea \prod_{i=1}^n U(1)^{N_i} \eea including the one corresponding to the trace
of $U(N_i)$.

At the Higgsed branch with the gauge group \bea \prod_{j=1}^n \prod_{k=j}^n \prod_{l=1}^{d_{j,k}} U(M_{j,k,l})\eea
the adjoints get the expectation values: \bea P_i(v) = \det (v- \Phi_i) = \prod_{j=1}^i \prod_{k=i}^n
\prod_{l=1}^{d_{j,k}}  (v-e_{j,k,l})^{M_{j,k,l}}.\eea In terms of
$U(1)^{N_i}$ theory which comes from the
factor $U(N_i)$, there are $N_i - \sum_{j=1}^i \sum_{k=i}^n d_{j,k}$ mutually local massless magnetic monopoles
when $N_i - \sum_{j=1}^i \sum_{k=i}^n d_{j,k}$ corresponding one
cycles collapse. Thus  the  massless photons can be described by the reduced curve
\bea \label{reducedN=2curve}u^{n+1} - \tilde{P}_1(v) u^n &+ \sum_{j=2}^{n} (-1)^j (\prod_{i=1}^{j -1}
\La_i^{(j-i)\be'_i})\tilde{P}_j(v) u^{n+1-j} \nonumber \\ &+ (-1)^{n+1}(\prod_{i=1}^{n} \La_i^{(n+1-i)\be'_i}) = 0, \eea where
\bea &\tilde{P}_i(v) = \prod_{j=1}^i \prod_{k=i}^n \prod_{l=1}^{d_{j,k}} (v-e_{j,k,l}),&\nonumber \\ &\be'_i = 2\sum_{j=1}^i
\sum_{k=i}^n d_{j,k}-\sum_{j=1}^{i-1} \sum_{k=i-1}^n d_{j,k}- \sum_{j=1}^{i +1}\sum_{k=i+1}^n d_{j,k}.& \eea The
reduced curve is an $(n+1)$-fold branched covering of the complex
plane whose branch cuts are centered around the
eigenvalues $e_{j,k,l}$ of $\Phi$. The eigenvalues of two different adjoints can be the same.  The number of
different eigenvalues of all the adjoints is  exactly $d$ which is given by (\ref{d}). After  the NS branes become
curved in the $u,~v$-space due to the expectation values of the adjoints and the bifundamentals, every different
eigenvalue $e_{j,k,l}$ will split into two points with ramification index two. Hence the Euler characteristic
$\chi$ of the reduced curve after curving have the following relation by Riemann-Hurwitz
formula: \bea \chi:= 2-2g = 2(n+1) -2d \eea so that the genus $g$ is equal to $d-n$ which is the same as
(\ref{gg}). When $\mbox{deg $\Phi_i$} =m$ for all $i$, then the genus is  \bea g =\frac{ m(n+1)n}{2} -n, \eea
which is equal to (\ref{Gg}). Moreover, the reduced curve after curving and translation is the same as
(\ref{SigmaNS}) as they describe the same field theory.

In fact, we can show more precisely how the parameters $\zeta_{i,j}$ of (\ref{GeomCurve}) are related to the
parameters and the $\CN=2$ QCD scales. For simplicity, we consider the $\A_2$ theory where the gauge group breaks
as \bea U(N_1) \times U(N_2) \rightarrow U(M_1)\times U(M_2) \times U(M_{12}) \eea The $\CN =2$ Seiberg-Witten
curve is written as:
 \bea \Lambda_1^{2N_1-N_2}t^3 -
&\hspace{-.5in}(v-a_1)^{M_1}(v-a_{12})^{M_{12}} t^2 \hspace{1in}\nonumber \\ &+ (v-a_2)^{M_2}(v-a_{12})^{M_{12}} t - \Lambda_2^{2N_2 -N_1} =0,\eea when the
D4 between the first and second NS branes are split into two stacks located at $v=a_1, a_{12}$ and D4 between
the second and the third at $v=a_2, a_{12}$. From (\ref{fterm}), $a_{12}$ is necessarily the center of the mass
coordinates i.e. $a_{12} = (a_1g_1 + a_2g_2)/(g_1+g_2)$. Then in the limit,
we have
\bea \bullet& v\to a_1,& (v-a_1)^{M_1} \sim \Lambda_1^{2N_1-N_2}t (a_1-a_{12})^{-M_{12}} + (a_1-a_2)^{M_2} t^{-1}\nonumber \\
&&+ \Lambda_2^{2N_2 -N_1} t^{-2}(a_1-a_{12})^{-M_{12}},\nonumber \\\bullet& v\to a_2,& (v-a_2)^{M_2} \sim -
\Lambda_1^{2N_1-N_2}t^2(a_2-a_{12})^{-M_{12}} \nonumber \\&&+ (a_2-a_1)^{M_1} t (a_2-a_{12})^{-M_{12}}\nonumber\\&&
+\Lambda_2^{2N_2 -N_1}t^{-1}(a_2-a_{12})^{-M_{12}},\nonumber \\ \bullet& v\to a_{12},&  (v-a_{12})^{M_{12}} \sim
\frac{ \Lambda_1^{2N_1-N_2}t^3 + \Lambda_2^{2N_2 -N_1}}{-
(a_{12}-a_1)^{M_1} t^2 + (a_{12}-a_2)^{M_2}t}. \label{N2approx}\eea From
the projection (\ref{GeomCurve}) of $\CN =1$ M-theory curve onto $u-v$ plane , we have \bea \label{A}u(v-a_i)\sim
(-1)^{i-1}\frac{\zeta (0, a_i)}{g_1g_2(a_1-a_2)}:= \mu_{ii}\eea around $u=0, v =a_i, i=1,2$. In order to match the
first two parameters $\mu_{11}$ and $\mu_{22}$, you restrict $\CN =1$
curve over the line $u=0$. Then we have \bea
t \sim
(v-a_1)^{-M_1}(v-a_2)^{M_2}. \eea So we have \bea\nonumber t \sim
(v-a_1)^{-M_1}(a_1-a_2)^{M_2}, \,\, \lim (u,v)=(0,a_1)\\
\label{B}  t \sim (a_2-a_1)^{-M_1}(v-a_2)^{M_2}, \,\, \lim (u,v)=(0,a_2).\eea The mass $g_i$ of $\Phi_i$
causes the $i$-th NS brane in $\CN=1$, on which $M_i$ of D4 brane end on the left, rotate to $u=g_i(v-a_i)$.
This together with (\ref{N2approx}) implies that \bea
g_1^{-M_1}u^{M_1} = (v-a_1)^{M_1} &\sim & \Lambda_1^{2N_1-N_2}t (a_1-a_{12})^{-M_{12}} + (a_1-a_2)^{M_2} t^{-1}\nonumber \\
&&+ \Lambda_2^{2N_2 -N_1} t^{-2}(a_1-a_{12})^{-M_{12}}\nonumber \\
g_2^{-M_2}u^{M_2} = (v-a_2)^{M_2} &\sim &- \Lambda_1^{2N_1-N_2}t^2(a_2-a_{12})^{-M_{12}} \nonumber\\&& + (a_2-a_1)^{M_1} t
(a_2-a_{12})^{-M_{12}}\nonumber\\&& -\Lambda_2^{2N_2
-N_1}t^{-1}(a_2-a_{12})^{-M_{12}} \label{C} \eea On the other hand, we
obtain from (\ref{A}) and (\ref{B}) that \bea \mu_{11}^{-M_1} u^{M_1} \sim t (a_1-a_2)^{-M_2}\nonumber\\
\label{N1approx} \mu_{22}^{-M_2} u^{M_2} \sim t^{-1} (a_2-a_1)^{-M_1}. \eea By comparing (\ref{C}) and
(\ref{N1approx}) for the large (resp. small) $t$ for $\mu_{11}$ (resp. $\mu_{22}$), we obtain
\bea \mu_{11}^{M_1} = g_1^{M_1} (a_1-a_{12})^{-M_{12}} (a_1-a_2)^{M_2} \Lambda_1^{2N_1-N_2} \\
\mu_{22}^{M_2} = g_2^{M_2} (a_1-a_{12})^{-M_{12}} (a_2-a_1)^{M_2} \Lambda_2^{2N_2-N_1}
 \eea
To compare the scales for the remaining gauge group $U(M_{1,2})$, we take
the following form of the $\CN=2$ curve
 \bea \label{NEW} t^3\!-\! (v\!-\!a_1)^{M_1}(v\!-\!a_{12})^{M_{12}}
t^2 \!+\! \La_1^{\be_1}(v\!-\!a_2)^{M_2}(v\!-\!a_{12})^{M_{12}} t
\!-\!\La_1^{2\be_1}\Lambda_2^{\be_2} \!=\!0,\eea which gives \bea\bullet& v\to
a_{12},& (v-a_{12})^{M_{12}} \sim \frac{ t^2 - \Lambda_1^{2\be_1} \La_2^{\be_2}t^{-1}}{- (a_{12}-a_1)^{M_1} t+
\La^{\be_1}(a_{12}-a_2)^{M_2}}.\eea

In order to match the deformation parameter corresponding to $U(M_{12})$, we consider $\CN =1$ curve along the
line $u=g_1(v-a_1)$.  First we will make coordinate changes so that the lines $u-g_1(v-a_1)$ and $u+g_2(v-a_2)$
move to the lines $u=0$ and $v-a_{12}$ and the points $(0, a_1)$ and $( g_1(a_{12} -a_1), a_{12})$ to $(0, a_1)$
and $(0, a_{12})$ in the new coordinate system. This can be done by the change $u -g_1(v-a_1) \to u $ and $(a_1
-a_{12})(u+g_2(v-a_2)) / (g_2(a_1 -a_2))+ a_{12} \to v $. Under the new coordinate system, the special direction
of $\CN=1$ curve is given by \bea t \sim (v-a_1)^{-M_1}(v-a_{12})^{M_{12}} \eea Thus $t \sim (a_{12} -a_1)^{-M_1}
(v-a_{12})^{M_{12}}$ around $(0, a_{12})$. We may approximate (\ref{SigmaNS}) by  \bea \label{a12curve}
u(v-a_{12})\sim \frac{ g_2(a_1-a_2)\z(a_{12}, g_1(a_{12} -a_1))}{a_{12}(a_1-a_{12})}:=\mu_{12}.\eea
 Since there will be mass contributions from both of the adjoins $\Phi_1, \Phi_2$, the curve
will rotate to $u= (g_1 + g_2)(v-a_{12})$ by adding superpotential. By comparison for small $t$, we obtain \bea
\mu_{12} = g_{12}^{M_{12}} (a_{12} -a_1)^{-M_1}(a_{12} -a_2)^{-M_2} \La^{\be_1} \La^{\be_2} \eea where $g_{12}
=g_1 +g_2$ and $\be_i$ is the beta-function for $U(N_i)$.

As in \cite{GUD}, the scales $\Lambda _{j,k}$ of the low energy $U(M_{j,k})$ theory can be determined by naive
threshold matching relations at the scales of all massive $U(M_{j,k})$ matter and W-boson fields: \bea
\label{scales}  \ljk^{3\mjk} = \gjk^\mjk 
\hspace{-.1in}
\prod_{(l,m) \in I^+}
\hspace{-.1in}
(\ajk - \alm)^{\mlm} 
\hspace{-.1in}
\prod_{(l,m) \in I^-}
\hspace{-.1in}
(\ajk -
\alm)^{-\mlm} \prod_{i=j}^k \li^{\be_i}\eea where $I^+ = \{(l,m)| m= j-1 \,\mbox{or}\,\, l=k+1 \}$ and $I^{-} = \{
(l,m) |l=j\,\, \mbox{or}\,\, m = k  \}- \{(j,k) \}.$ Recall that we always assume that the first index is less
than equal to the second index in the double indices. For the case under consideration, $(\ref{scales})$ can be
written as \bea \La_{1,1}^{3M_1} = g_1^{M_1} (a_1 - a_{1,2})^{-M_{1,2}} (a_1 -a_2)^{M_2} \La_1^{2M_1 - M_2
+M_{1,2}}, \nonumber \\
\La_{2,2}^{3M_2} = g_2^{M_2} (a_2 - a_{1,2})^{-M_{1,2}} (a_2 -a_1)^{M_1} \La_2^{2M_2 - M_1
+M_{1,2}}, \\
\La_{1,2}^{3M_{1,2}} = g_{1,2}^{M_{1,2}} (a_{1,2} - a_1)^{-M_{1}} (a_3-a_2)^{-M_2} \La_1^{2M_1 - M_2
+M_{1,2}}\La_2^{2M_2 - M_1 +M_{1,2}}. \nonumber  \eea Therefore, we can identify the geometric deformation parameter
$\mu_{ij}$ with $\CN =1$ QCD scale $\Lambda_{M_{i,j}}^3$ of $U(M_i)$.

\section{Orientifold Theories}
The whole discussions can be extended to the $SO/Sp$ gauge theories as in \cite{eot, dot2} in the presence of an
orientifold O6 plane. The A-D-E singularities defined by (\ref{ADEsingularities}) are invariant under the complex
conjugation on the ambient space $\C^3$ \bea (x,y, u) \to (\bar{x}, \bar{y}, \bar{u}).\eea  Moreover, this action
can be extended to the resolved ALE space where the exceptional $\P^1$ becomes $\RP^2$. By wrapping $N_i$ D5
branes on each $\P^1$ we obtain the $\CN =2$ supersymmetric gauge theories with gauge group \bea \prod_{i=1}^n
O(N_i) \eea from type IIB theory with orientifolding.

Now consider the $\CN =1$ supersymmetric gauge theory deformed by a tree-level superpotential of the form: \bea
\label{sosupotg} W_{SO} =\sum_{i=1}^{n} W_i -\mbox{Tr}\sum_{i=1}^{n}\sum_{j=1}^n s_{i,j}Q_{i,j} \Phi_{j}
{Q}_{j,i},\mbox{where }W_i = \mbox{Tr}\sum_{j=1}^{d_i+1} \frac{g_{i,2j-1}}{2j} \Phi_i^{2j}. \eea We can obtain
the classical vacua using the method used in Section 4. The eigenvalues of the adjoints are the roots of \bea W_j'
(v) + W_{j+1}'(v) + \ldots + W_k'(v) = g_{2d_{j,k}} v \prod_{l=1}^{d_{j,k}} (v^2 + b_{j,k,l}^2),~~b_{j,k,l} > 0,
\eea where $g_{2d_{j,k}}$ is the sum of the highest coefficient of $W_i$ whose degree is maximal among $W_j,
\ldots, W_k$. Following the brane interpretation considered before,  we conclude that the $\CN=1$ geometry is
given by the resolution of \bea xy - u\prod_{p=1}^n\left( u - \sum_{i=1}^p W_i'(v) \right) =0. \eea Here the
singularities are arranged so that the geometry is invariant under the orientifold action \bea (x,y,u,v) \to (
\bar{x}, \bar{y}, \bar{u}, \bar{v}). \eea Because of the presence of the orientifold plane located at $u=v=0$, the
$\P^1$ cycle at $u=v=0$ becomes an $\RP^2$ cycle stuck on the orientifold plane and the gauge theory on the D5
brane wrapped on it is $O(2M_0)$. The D5 branes on $\P^1$ located at $v= ib_{j,k,l}$ are identified with the D5
brane on $\P^1$ located at $v=-ib_{j,k,l}$ and the gauge theory on the corresponding D5 brane is $U(M_{j,k,l})$.
Therefore the gauge group will break: \bea \prod_{i=1}^n O(N_i) \to \prod_{i=1}^n
O(M_i)\prod_{j=1}^n\prod_{k=j}^n\prod_{l=1}^{d_{j,k}} U(M_{j,k,l}), \eea where $d_{j,k}$ are defined in
(\ref{djk}). Under geometric transition, the $\P^1$ cycles shrink and is replaced by $\S^3$. The $\S^3$ located at
$u=v=0$ is invariant under the orientifold action, and the $\S^3$ located at $v= ib_{j,k,l}$ maps to one located
at $v =- b_{j,k,l}$ and vice-versa.

Since the circle actions are compatible with the orientifolding, we can take T-dual pictures and lift to M-theory.
By M-theory transition, we obtain a Seiberg-Witten curve in the $u-v$ plane for $\CN=1$ $U(1)^d$ theory with
$\Z_2^n$ global symmetries generalizing the results of \cite{eot, dot2}.

One can also use orientifolds O4 planes which would imply that we start with an  $\CN=2$ theory with product group
\bea O(N_1) \times Sp(N_2) \times O(N_3) \times Sp(N_4) \times \cdots \eea This can be again deformed to an
$\CN=2$ supersymmetric theory and the result is similar to the ones obtained before and are a generalization of
the ones of \cite{rt}.

\section{Matter Fields and Seiberg Dualities}
We can also discuss the Seiberg duality \cite{seiberg} which takes
place in the presence of quark chiral superfields in the fundamental
representation of \bea
\prod_{j=1}^n\prod_{k=j}^n\prod_{l=1}^{d_{j,k}} U(M_{j,k,l})\eea in
$\CN=1$ theory, by generalizing the results of \cite{dot2} (for related discussions
concerning connections between Seiberg duality and toric dualities see
\cite{plesser,GUD}). Here we consider the
case of massive fundamental fields, which will be integrated out in
the infrared which changes the scale of the $\CN=1$ theory so changes
the flux on the $\S^3$ cycles.

As in \cite{civ, dot2}, the massive matter fields can be obtained by wrapping D5 branes on non-compact holomorphic
curves at distance from the $\P^1$ cycles and the distances are identified with mass of the hypermultiplet. Recall
that the $\CN=1$ geometry is covered by open sets $U_p, ~~p=0, \ldots, n$ whose coordinates are given by $u_p,
x_p, v_p$. The holomorphic curve  defined by \bea\label{matter}  v_p =m ,~~ x_p =0 \eea in $U_p$ maps to $x=0, u=
\sum_{i=0}^p W'_i(m), v=m$ under the blown-down map $\sigma$ (\ref{sigma}). Hence if $v=m$ does not pass through
one of the intersection points of the NS branes in the $u-v$ plane, then the holomorphic curve (\ref{matter})
cannot be  compactified in $\tilde{X}$  and D5 branes wrapped on the curve become semi-infinite D4 branes bounded
by a NS brane on one side.

The  $\CN=1$ Seiberg dualities for these models are generalizations
for the results of \cite{kuta} to the case of product gauge group.
In order to describe the  $\CN=1$ Seiberg dualities one starts with
the  $\CN=2$ Seiberg dualities discussed in \cite{9603042}. The $\CN=1$ theories with fundamental matter
will have a classical moduli space given by the solution of the D-term
and F-term equations which for the case of  \cite{9603042}
imply different branches, the Coulomb branch (when the vacuum
expectation values of the fundamental fields
are zero) and the Higgs branch (when the vacuum expectation values of the
adjoint field is zero), the latter being split into a non-baryonic
branch and a baryonic branch. By adding a superpotential for the
adjoint fields, the results of \cite{9603042} imply that the baryonic
branch and some isolated points on the non-baryonic branch are non
lifted and they are points where $\CN=1$ supersymmetry is preserved.
It has been observed that, in the $\CN=2$ theories, points in the
baryonic branch have a double description, either as classical
solutions for an $\CN=2$ $U(N)$ theory with $F$ fundamental flavors or
for an $\CN=2$ $U(F-N)$ theory with $F$ fundamental flavors and these two
theories are Seiberg dual to each other. We can break the
supersymmetry in both theories  to   $\CN=1$ theory
by adding a superpotential for the adjoint fields.
If we denote the adjoint field for the  $U(N)$ theory by $\Phi$
and the one for the  $U(F-N)$ theory by $\tilde{\Phi}$, the
corresponding $\CN=1$ theories are obtained by adding the
superpotentials
\bea
W(\Phi) = \sum_{k>0} \mbox{Tr} \Phi^k
\eea
or
\bea
W(\tilde{\Phi}) = \sum_{k>0} \mbox{Tr} \tilde{\Phi}^k
\eea
respectively. Moreover, the two corresponding $\CN=1$ theories are
Seiberg dual to each other. To give a geometrical picture,
we start in the brane side of the type IIB picture where the matter
field is added as D5 branes wrapping holomorphic 2-cycles separated
by a distance $m$ from the exceptional $\P^1$.

As discussed in
\cite{dot2}, the Seiberg duality is seen in geometry as a birational flop
transition where the exceptional $\P^1$  cycles are replaced by
another $\P^1$ cycles with negative volume. From the above discussion,
we conclude that all the exceptional
$\P^1$
cycles in the  $\CN=1$ theory are replaced together in a single flop
because all of them originate  from one $\P^1$
cycle in $\CN=2$ theory after the superpotential deformation.
In type IIA picture, this implies that one have to move one NS brane
through another NS brane completely and one cannot move just some part of
NS brane, while keeping other part fixed,  through another NS brane as all of D4 branes stretched between them
change together. In terms of geometry, one possible way of achieving
the Seiberg-duality by changing the coordinates of $U_k$ (\ref{sigma}):
\bea
u_p = \frac{\prod_{j=0}^{p}\left(u- \sum_{i=0}^j W_i'(v)\right)}{x},x_{p} = \frac{x}{\prod_{j=0}^{p-1}\left(u-
\sum_{i=0}^j W_i'(v)\right)}, v_p =v,\eea
to \bea
\tilde{u_p} = \frac{\prod_{j=p}^{n}\left(u- \sum_{i=0}^j W_i'(v)\right)}{x}, \tilde{x_{p}} = \frac{x}{\prod_{j=p+1}^{n}\left(u-
\sum_{i=0}^j W_i'(v)\right)}, v_p =v.\eea

Hence we conclude that we cannot randomly close $\P^1$
cycles but we need to close all the $\P^1$ cycles corresponding to the
intersection points of two curved NS branes. This is related to the
fact that the deformations of the complex structure for the dual
theories are related.

To clarify the above claim, we consider the example of the ${\bf A_3}$ theory
as in  Figure 6. We have 17 gauge
groups in the  ${\cal N} = 1$ theory for which we could discuss the
Seiberg duality. But the electric and magnetic theory should come
from Seiberg dual ${\cal N} = 2$ theories with the corresponding
superpotential deformations. As discussed before, this means that we
have to close groups of $\P^1$ cycles which correspond to all the
intersection
points between two NS curved branes. In the previous section
notations, this means that we need to close the $\P^1$ cycles
corresponding to groups of points $(p_{011}, p_{012})$ (which
correspond to the intersections between the 0-th NS brane and the 1-st
NS brane), $(p_{021}, p_{022}, p_{023})$  (which
correspond to the intersections between the 0-th NS brane and the 2-nd
NS brane), $(p_{031}, p_{032}, p_{033})$ (intersections 0-th and 3-rd),
$(p_{121}, p_{122}, p_{123})$ (intersections 1-st and 2-nd),
$(p_{131}, p_{132}, p_{133})$ (intersections 1-st and 3-rd),
$(p_{231}, p_{232}, p_{233})$ (intersections 2-nd and 3-rd). In terms
of the gauge groups, this means that we have to consider the Seiberg
dual for the product group
$U(M_{121}) \times U(M_{122}) \times U(M_{123})$ and other product groups
read from the previous intersection points.

This is an interesting and
somehow unexpected feature of the Seiberg dualities for the product
groups with bifundamentals. In order to understand it better, we need
to know the moduli space for these theories in the
presence of fundamental matter, which involves a discussion similar to
the one in the previous sections. One thing appears in the type IIA
brane configuration related to the positions of the semi-infinite D4
branes corresponding to massive flavors. In the case of non-curved NS
branes, the positions of the  semi-infinite D4 branes are related to
the masses of the fundamental flavors but in the case of curved NS
branes this is more difficult to see. The semi-infinite D4 branes
will be distributed around the intersection points and so we have a
flavor symmetry breaking in the ${\cal N} = 1$ theory. This might be
related to the above described phenomenon of $\P^1$ cycles closing and
opening in groups. We hope to return to this issue in the future.

\section{Discussion}

One important question is whether we can compare our type IIA brane
configuration construction with other type IIA
constructions as the ones in \cite{av,av1}. This would relate our MQCD
with M-theory curves to M-theory with  $G_2$ manifolds. Moreover, one may
be able to construct the various $G_2$ holonomy manifolds
which have been used in M-theory lifts of the geometrical type IIA transitions \cite{achar} (see other
important developments in \cite{achar1}-\cite{lust}).

The type IIA and type IIB pictures are related by mirror symmetry
which was interpreted as three T-dualities in \cite{syz}. In toric
geometry, the natural T-dualites come from $U(1)^3$ of the torus embedding.
In fact, the case for the conifold and its Abelian quotients have been
studied in  \cite{miemiec}.  Recently, geometric transitions in the toric
situation has been studied in \cite{av1}  utilizing the mirror symmetries of
\cite{hori1, hori2}.

Our type IIA picture has been obtained by taking
 one T-duality  from type IIB picture. Hence we need to take two more
 T-dualities to obtain the mirror type IIA.
It would be interesting to work out in details for the toric cases as in
 \cite{av1} to see how two type IIA picture appear and how their M-theory
 transitions are related. More generally, the geometry is not toric and
 there are no obvious three T-dualities one can take.
In the case of degenerate superpotential with the same degree,
the geometry (\ref{singandegd})  become quasi-homogeneous and there is
 extra $U(1)$ action.  Brane
configurations have been extensively studied during the last years and
in the present paper we extend them to more general cases. Therefore,
after two extra T-dualities we could get a rich class of $G_2$ manifolds.
\section{Acknowledgements}

We would like to thank  Mina Aganagic, Ami Hanany, Kentaro Hori, Keshav Dasgupta, Ansar Fayazzudin, Sergei Gukov,
Albrecht Klemm, Karl Landsteiner, Anatoly Libgober, Cumrun Vafa for insightful and valuable discussions. We would
also like to thank BumHoon Lee and Soonkeon Nam for their participation in the early stage of this work. We would
like to thank Cumrun Vafa and Harvard University for invitation and hospitality.

\end{document}